\documentclass[
reprint,
superscriptaddress,
showpacs,preprintnumbers,
bibnotes,
 amsmath,amssymb,
 aps,
 prl,
floatfix,
]{revtex4-1} 

\usepackage{amssymb,bm}
\usepackage{amsmath}
\usepackage{amsthm}
\usepackage{graphicx}
\usepackage{hyperref}
\hypersetup{pdfborder=0 0 0}

\bmdefine\Bone{1}

\newcommand{\Trc}{\mathrm{Tr}\,}

\newcommand{\hf}{\displaystyle\frac{1}{2}}

\newcommand{\curl}{\mathrm{curl}\,}

\newcommand{\dif}[2]{\displaystyle\frac{\partial #1}{\partial #2}}
\newcommand{\Grad}{\nabla}

\newcommand{\Md}{\partial}

\newcommand{\Tld}[1]{\widetilde{#1}}

\def\XXint#1#2#3{{\setbox0=\hbox{$#1{#2#3}{\int}$ }
\vcenter{\hbox{$#2#3$ }}\kern-.6\wd0}}

\newcommand{\jump}[1]{\lbrack\!\lbrack #1 \rbrack\!\rbrack}
\newcommand{\lump}[1]{\lbrace\skew{-14.7}\lbrace\!\!#1\!\!\skew{14.7}\rbrace\rbrace}

\newcommand{\mat}[4]{\left[\begin{array}{cc}
\displaystyle{#1}&\displaystyle{#2}\\[1ex]
\displaystyle{#3}&\displaystyle{#4}\end{array}\right]}

\newcommand{\lhs}{left-hand side}

\newcommand{\nbh}{neighborhood}

\newcommand{\Ga}{\alpha}
\newcommand{\Gb}{\beta}

\newcommand{\Ge}{\epsilon}

\newcommand{\Gg}{\gamma}

\newcommand{\Gl}{\lambda}

\newcommand{\Gth}{\theta}

\newcommand{\GD}{\Delta}

\newcommand{\GL}{\Lambda}

\bmdefine\BGa{\alpha}
\bmdefine\BGb{\beta}
\bmdefine\BGd{\delta}
\bmdefine\BGe{\epsilon}
\bmdefine\BGve{\varepsilon}
\bmdefine\BGf{\phi}
\bmdefine\BGvf{\varphi}
\bmdefine\BGg{\gamma}
\bmdefine\BGc{\chi}
\bmdefine\BGi{\iota}
\bmdefine\BGk{\kappa}
\bmdefine\BGl{\lambda}
\bmdefine\BGn{\eta}
\bmdefine\BGm{\mu}
\bmdefine\BGv{\nu}
\bmdefine\BGp{\pi}
\bmdefine\BGth{\theta}
\bmdefine\BGvth{\vartheta}
\bmdefine\BGr{\rho}
\bmdefine\BGvr{\varrho}
\bmdefine\BGs{\sigma}
\bmdefine\BGvs{\varsigma}
\bmdefine\BGt{\tau}
\bmdefine\BGj{\tau}
\bmdefine\BGu{\upsilon}
\bmdefine\BGo{\omega}
\bmdefine\BGx{\xi}
\bmdefine\BGy{\psi}
\bmdefine\BGz{\zeta}
\bmdefine\BGD{\Delta}
\bmdefine\BGF{\Phi}
\bmdefine\BGG{\Gamma}
\bmdefine\BGL{\Lambda}
\bmdefine\BGP{\Pi}
\bmdefine\BGT{\Theta}
\bmdefine\BGS{\Sigma}
\bmdefine\BGU{\Upsilon}
\bmdefine\BGO{\Omega}
\bmdefine\BGX{\Xi}
\bmdefine\BGY{\Psi}

\bmdefine\BFM{\mathfrak{M}}
\bmdefine\BFb{\mathfrak{b}}
\bmdefine\BFk{\mathfrak{k}}
\bmdefine\BFm{\mathfrak{m}}
\bmdefine\BFu{\mathfrak{u}}
\bmdefine\BFv{\mathfrak{v}}

\newcommand{\CB}{{\mathcal B}}

\newcommand{\CJ}{{\mathcal J}}

\newcommand{\CS}{{\mathcal S}}

\newcommand{\CW}{{\mathcal W}}

\bmdefine\BCA{{\mathcal A}}
\bmdefine\BCB{{\mathcal B}}
\bmdefine\BCC{{\mathcal C}}
\bmdefine\BCD{{\mathcal D}}
\bmdefine\BCE{{\mathcal E}}
\bmdefine\BCF{{\mathcal F}}
\bmdefine\BCG{{\mathcal G}}
\bmdefine\BCH{{\mathcal H}}
\bmdefine\BCI{{\mathcal I}}
\bmdefine\BCJ{{\mathcal J}}
\bmdefine\BCK{{\mathcal K}}
\bmdefine\BCL{{\mathcal L}}
\bmdefine\BCM{{\mathcal M}}
\bmdefine\BCN{{\mathcal N}}
\bmdefine\BCO{{\mathcal O}}
\bmdefine\BCP{{\mathcal P}}
\bmdefine\BCQ{{\mathcal Q}}
\bmdefine\BCR{{\mathcal R}}
\bmdefine\BCS{{\mathcal S}}
\bmdefine\BCT{{\mathcal T}}
\bmdefine\BCU{{\mathcal U}}
\bmdefine\BCV{{\mathcal V}}
\bmdefine\BCW{{\mathcal W}}
\bmdefine\BCX{{\mathcal X}}
\bmdefine\BCY{{\mathcal Y}}
\bmdefine\BCZ{{\mathcal Z}}

\bmdefine\Bzr{ 0}
\bmdefine\Ba{ a}
\bmdefine\Bb{ b}
\bmdefine\Bc{ c}
\bmdefine\Bd{ d}
\bmdefine\Be{ e}
\bmdefine\Bf{ f}
\bmdefine\Bg{ g}
\bmdefine\Bh{ h}
\bmdefine\Bi{ i}
\bmdefine\Bj{ j}
\bmdefine\Bk{ k}
\bmdefine\Bl{ l}
\bmdefine\Bm{ m}
\bmdefine\Bn{ n}
\bmdefine\Bo{ o}
\bmdefine\Bp{ p}
\bmdefine\Bq{ q}
\bmdefine\Br{ r}
\bmdefine\Bs{ s}
\bmdefine\Bt{ t}
\bmdefine\Bu{ u}
\bmdefine\Bv{ v}
\bmdefine\Bw{ w}
\bmdefine\Bx{ x}
\bmdefine\By{ y}
\bmdefine\Bz{ z}
\bmdefine\BA{ A}
\bmdefine\BB{ B}
\bmdefine\BC{ C}
\bmdefine\BD{ D}
\bmdefine\BE{ E}
\bmdefine\BF{ F}
\bmdefine\BG{ G}
\bmdefine\BH{ H}
\bmdefine\BI{ I}
\bmdefine\BJ{ J}
\bmdefine\BK{ K}
\bmdefine\BL{ L}
\bmdefine\BM{ M}
\bmdefine\BN{ N}
\bmdefine\BO{ O}
\bmdefine\BP{ P}
\bmdefine\BQ{ Q}
\bmdefine\BR{ R}
\bmdefine\BS{ S}
\bmdefine\BT{ T}
\bmdefine\BU{ U}
\bmdefine\BV{ V}
\bmdefine\BW{ W}
\bmdefine\BX{ X}
\bmdefine\BY{ Y}
\bmdefine\BZ{ Z}

\begin{document}

\title{Rigidity-induced critical points} 

 \author{Y. Grabovsky}
 \affiliation{\it  Department of Mathematics, Temple University, Philadelphia, PA 19122, USA }

\author{L. Truskinovsky}
 \affiliation{\it  PMMH, CNRS-UMR 7636, ESPCI, PSL,  75005 Paris, France}

\begin{abstract}
While classical  theory of  phase transitions deals with  systems where  shape variation   is  energetically  neutral,   the account of  \emph{rigidity}  can lead to  the emergence  of  new thermodynamic features.  One of them is a special  type of critical points that are characteristic of  phase transitions specifically in solids.   We develop a general theory of  such  rigidity-induced critical points and  illustrate  the  results by analyzing  in detail the case of   an isotropic, geometrically nonlinear  solid undergoing a volumetric phase transition at zero temperature.
\end{abstract}

\date{\today}

\maketitle

 \section{ 1. Introduction}
  The conventional  thermodynamic approach to   phase stability of  solids is to treat them   mechanically  as liquids. 
This classical  perspective  has been   challenged by the appearance  of
'kinetic' or 'coherent' phase diagrams for diffusionless phase transitions
\cite{cahn62,will80,kms88,wozu89,akat11,coba12,fwz04,maug19,maug22,llsb23}
  and the attempts to understand the ensuing complex multiphase microstructures  dominated by elastic
  interactions \cite{kh,bhat03}.  To corroborate the idea  that  the account of rigidity in
  solid-solid transformations can lead to fundamentally new thermodynamic
  effects,  we focus in this paper on the emergence  in elastically compatible  systems of a peculiar type of critical points which do not exist in rigidity-free (liquid) systems. 
  
  \subsection{Motivation: swelling transition  in  gels} A particularly striking example of the failure of  the   'liquid'
   perspective on phase equilibria was provided by  experimental studies of
   polymeric gels exhibiting   swelling   phase transition
   \cite{lita92,onuki05}.    While the volumetric   transformation between the
   swollen and shrunken phases   resembles     a first-order  
   liquid-gas  transition,     several  peculiar 'solid' features observed during such transitions pointed towards the importance of nonzero shear  rigidity  
  \cite{tana86,hwka88,seka88,mata88,mata92}.  
     
Thus,  in  contradiction
   with classical thermodynamics of  such symmetry preserving isotropic-to-isotropic transitions    \cite{ruel69}, experiments  have shown  that a discontinuous   swelling of  a gel  can generate   inhomogeneous  patterns of  anisotropically stressed coexisting  phases.    Specifically,  experiments pointed towards   the formation of microstructures which are neither isotropic nor homogeneous  with developing    patterns   that do not resemble the ones controlled by liquid surface tension  \cite{tfsnss80,ssk89,onuk88}.
Moreover,  the implied domain microstructures were observed in the range of parameters
where, according to the 'liquid' approach,  they should have been  mechanically unstable 
 \cite{taii77,tfsnss80,onuk88,ssk89}. Independently,   experiments   showed the possibility of small, negative values of the compressibility  which, as an equilibrium effect, is a feature of nonzero rigidity. This, together with the fact 
  that critical fluctuations in  swelling  gels are not seen at zero bulk modulus,  as purely liquid thermodynamics would have predicted, suggests  that gels cannot be adequately  modeled by conventional thermodynamic theory  \cite{taii77,lakes2008negative}.

In the context of the transition between swollen and de-swollen gels, the   'non-liquid' effects have been previously  linked to  long-range interactions   pointing towards   mean-field type description \cite{larkin1969phase,golubovic1989nonlinear}.
It was also shown    that   due to   rigidity-induced  nonlocality,   mediated by transverse phonons,    gels can be  stabilized at negative values of the compressibility in the presence of sufficiently strong  boundary constraints. It was similarly argued  that  the specificity of swelling phase transitions in gels is 
  due to the absence of characteristic length in elasticity theory
  \cite{flor53}, which ensures that the  
  activation energy of nucleation in the bulk is proportional  to the
  volume   and is therefore macroscopic in contrast to what is usually postulated in classical thermodynamic theory 
  \cite{onuki1988paradox}.  
 
Building upon these insights, we intend to  shed additional light on  why
rigidity of gels and similar soft solids has such  a profound effect on their
equilibrium thermodynamics.  In particular, we'll be concerned with the fact
that in swelling gels the observations of critical opalescence remain
controversial, since the expected  'liquid' scaling response has not been
observed  at critical points, given that they are interpreted in the framework of classical thermodynamics \cite{tana78,tfsnss80}.

 \subsection{Coherency constraint} The crucial reason  for  developing  a purely   'solid' perspective on  diffusionless transformations in swelling gels is    that   polymer networks  apparently do  not change  their  connectivity during   swelling transitions. More generally, one can say  that behind the   nonzero value of  shear modulus  in polymer gels  is  the immutable network topology, and that it is the presence of an implied   topological   constraint  that  enables gels to resist elastic distortions.   Therefore,   the observed  deformations should be viewed as elastically 'coherent' in the sense that   the underlying  macroscopic displacement fields  are continuous and the associated deformation gradients are geometrically compatible in the sense of Hadamard \cite{seka87,onuk88a,seka89,onuk89,ssk90,para96,bsl96,mtmtt99}.   

To abstract  this idea, the  'coherent' thermodynamics  views a  solid  as always equipped with  a  fixed  reference state \cite{laca73,laca78,penr02,wiev09,wiev10,will11,kolu2014,saha18,nghs18}. Such an assumption is not straightforward  since   in principle  atoms can exchange places which   excludes the possibility  of  a nonlocal coherency  constraint  and  implies that the equilibrium value of the shear modulus must be zero  \cite{sbk10,penr02}. The emergence of rigidity is therefore ultimately associated with inherently long-living metastability \cite{saw2016rigidity}. 

In contrast,  the classical nonlinear elasticity theory, e.g.  \cite{podi13,green1992theoretical,ogdenbk97,lurie2012non,stoker1968nonlinear,ciarlet2021mathematical},
 ignores the  metastability aspect of elastic equilibrium  
 and views solids as 
 never losing topological memory about their local environments.  
In other words,  it postulates  the existence of 
  a topologically constrained  configuration space which, in particular, prohibits the mutual exchanges  of atomic positions  \cite{laca73,kolu2014,saha18,nghs18}.
 While  such a constraint is just an approximation,    at small temperatures   the  flow of defects which is supposed to relax the internal stresses   is    anomalously slow 
because the associated  effective viscosity diverges at vanishing shear stress with an essential singularity \cite{sbk10,agoritsas2016driven}. Therefore,  in normal conditions the    'kinetic'  phase diagrams,   accounting for 'metastable' rigidity are  usually  fully adequate 
 \cite{pome02,bdbdm17}.

The  presence of the configurational  constraint, requiring that a
unique reference state exists,  opens a possibility for a configuration of a 
  loaded  solid body  to be  nontrivial    at equilibrium.   Suppose that 
$\Bx$  is
  the position of a material point in the reference state and   $\By(\Bx)$  is   its position in the deformed state. The fundamental assumption of the theory of elasticity is that the   total  energy  can be written as an integral over the region occupied by the   body  in the reference configuration with the macroscopic  stored-energy density function  $W$  depending only on the deformation gradient
   \begin{equation} \label{grad}
  \BF= \nabla \By. 
  \end{equation}  
This assumption implies that  at zero temperature
 the deformed equilibrium   state    can be   found by minimizing the total energy
 \begin{equation} \label{grad1}
 \min_{\By(\Bx)} \int W( \BF) d \Bx,
 \end{equation}
   subject to boundary conditions which, in the case of hard device type loading,  would  involve restrictions on the boundary values of $\By(\Bx)$ \cite{ogdenbk97,gfa10,podi13, silh13, krro19}. 
   The complexity of the problem \eqref{grad1} is due to the fact 
 that the tensorial argument of the function $W$, which plays a role of  an  order parameter,  is  a gradient of a continuous function.   Therefore,  the deformation gradient $\BF$ at different points  cannot be varied independently \cite{penr02} and in addition to the Euler-Lagrange equations 
\begin{equation} \label{grad0} 
 \nabla\cdot \BP=0,
  \end{equation}
where $P_{i}^{\Ga} =\dif{W}{F_{\Ga}^{i}}(\BF )$ is the Piola stress tensor, it must satisfy a   nonlocal  coherency  constraint 
\begin{equation} \label{grad2} 
\curl\BF=0.
\end{equation}
The latter, however, is only relevant
in the presence of rigidity, as for liquids
the constraint 
\eqref{grad2}   is  inactive  and the effective locality of the
minimization problem is recovered. It is also clear that,  in case of
generic loading, the solutions of \eqref{grad1}, satisfying \eqref{grad0}, \eqref{grad2}, may be highly  inhomogeneous.
   
  Under the assumption  of either   long living metastable configurations or  internally constrained equilibrium, it is  meaningful to develop  the ’coherent’ thermodynamics of solid-solid phase transitions which now counts many important contributions, e.g.  \cite{kh,laca84,roit84,jovo87}. 
This  and other related work has already identified  some  generic thermodynamic anomalies associated with  elastic   phase transitions  by linking them to   long-range elastic interactions induced by the coherency constraint \eqref{grad2}, see, for instance, \cite{lapi69,sak74,golu89,kh}. 
In particular, it was understood that  in  systems with nonzero rigidity  the   energy of phase mixtures depends not only on the volume fractions 
 but also on the detailed microstructure of coexisting phases including both the shape and the  orientation of the single phase domains  \cite{shkh95,shkh06,bhat03}.   
 
  Moreover, it was shown  that if the bottoms of the energy wells
  are not geometrically compatible,  in the sense that the
  corresponding values of  $\BF$ are not rank-one-connected,   mixing
  would  have  an extensive   energy cost \cite{DW,baja15}. An
  important resulting  effect  is  the  nonlinearity of the
  dependence of the elastic energy on the volume fractions of the
  phases in coherent multi-phase mixtures  \cite{shkh95,shkh06}. The
  implied   non-additivity of the   energy is also behind  the
  macroscopic energy barrier for phase nucleation which  is
  responsible for  metastability and rate independent hysteresis
accompanying  coherent  transformations \cite{baja15,jwk19}. All
  this invalidates   the use of the common tangent (Maxwell) construction  for
  determining phase equilibria   \cite{cahn62,will80,laca84,will84,john87},   replacing the usual convexification of the energy with a more subtle construction known as  quasi-convexification \cite{silh13,krro19,morr52,ball7677,daco82}.
 
 In case of gels undergoing swelling phase transitions  the above effects can be expected to be  manifestly  present. Thus,  since in the case of volumetric transitions the bottoms of the energy wells are not geometrically compatible,   the two-phase polymer network must deform inhomogeneously. Also, the observations of negative compressibility at such transitions in a range of parameters are supported by the fact that, in sufficiently constraining  loading conditions, the elastic instability, associated with volumetric phase transitions, takes place  strictly after the convexity of the energy has already been violated \cite{erto56,grtrmms}.

 \subsection{Organization of  the paper}  
 
 One of  the most important  signature of  'coherent' thermodynamics is that the tensorial stress becomes a parameter on phase diagrams replacing the conventional  liquid  pressure  \cite{laca84,roit84,jovo87,ftz03}. Our goal is to corroborate the idea  that such an  extension of the parameter space produces
new  qualitative effects. In particular, we intend to show  that in isotropic solids, in addition to the classical 'liquid' critical points associated with volumetric phase transitions, one can also expect the emergence of specific,  zero temperature, rigidity-induced  purely 'solid-type'  critical points.  

Critical points in the configurational space can be anticipated if there is no
distinction between the transforming phases in terms of their symmetry as they
can be in principle continuously deformed one into another.  To illustrate
this general idea we study in this paper as an example a class of phase
transformations between isotropic solid phases characterized by different
reference densities.  The phases are supposed to be subjected to homogeneous
(affine) deformations applied on the boundary.  In the course of such a loading
program, the deformation gradient $\BF$ is expected to cross into the coherent
binodal region 
\cite{grtrmms,grtrpcx}, where the system loses stability against strong
perturbations i.e., the infinitesimal perturbations of the deformation
$\By(\Bx)$ itself that are not infinitesimal as far as the deformation
gradient $\nabla \By (\Bx)$ is concerned. The boundary of the coherent binodal
region (also known as a \emph{ coherent binodal}) is then a set in the phase space
of tensors $\BF$ consisting of points where the energy density and its
quasi-convex envelope separate \cite{grtrpe,grtrmms,grtrpcx,grtrnc}.  This
makes the corresponding homogeneous configurations marginally stable, while
announcing the formation of an
inhomogeneous energy-minimizing microstructures immediately upon crossing the coherent binodal.  In typical cases, the coherent
binodal region can be expected to separate two (or more) connected components
of the phase space, which can be then identified as phases. This is, for
instance the case for liquids where the energy density $W$ depends only on the
specific volume $\det(\BF)$.  However, when rigidity is different from zero,
the stable part of the phase space can form a connected set cutting through the domain  of phase coexistence and permitting passage
from one phase to the other without any sharp transition and, consequently,
without the concomitant microstructure development.

In such cases one can define coherent critical points as the common limit of the two coexisting deformation gradients in
  two-phase equilibria, i.e. points at which   distinction between the two  phases in equilibrium
  disappears.   The emergence of such  points corresponds to the
  limiting case  when the range of phase coexistence and the attendant  microstructure
  development can shrink  to a point.  As in the case of $P-V$ diagrams for liquid phase transitions, in such critical points coherent
binodal and coherent spinodal in the space of deformation gradients $\BF$ are
tangent to each other,  which opens the way towards analytical characterization
of coherent critical points. Here by \emph{coherent spinodal} we designate the boundary of
elastic stability against local weak perturbations, i.e., infinitesimally
small perturbations of the deformation gradient localized in an infinitesimal
\nbh\ of a material point \cite{gagne2005simulations,herrmann1982spinodals,
  grtrmms}.

The  goal of this paper is to elucidate   these ideas,  support them by explicit analytical computations,   and then illustrate them  through a systematic  development of  an  example directly applicable to the description of swelling phase transitions in gels.
 
 We begin the paper in Section 2 by recalling various mechanical aspects of  the classical   first order symmetry preserving phase transition in a Van der Waals type liquid phase. While the analysis is performed using nonlinear elasticity theory,  it is  assumed that in equilibrium all non-hydrostatic stresses relax and   the elastic response can be fully represented by pressure-density relation. In the case of interest, the system   is   below  the classical liquid criticality,  moreover,  for simplicity, we  assume that the temperature is equal to zero. Under these conditions we determine the domain of equilibrium phase coexistence which is bounded by the classical thermodynamic binodal;  we also determine the location of the classical thermodynamic  spinodal and show that the two neither touch nor intersect. We then compute the equilibrium  energy parametrized by the specific volume which describes  the ground state. Expectedly, in this context, the  non-classical coherent critical points do not appear.

After reviewing these classical results, we present in Section 3 a
parallel  analysis for the model of a  generic diffusionless phase
transition in nonlinear elastic solid,  exhibiting general rigidity
and allowing for arbitrary transformation strain. In this way we
permit  the structure of the binodal to change  qualitatively vis a
vis the case of an elastic liquid, and open the way towards the
possibility of non-classical, rigidity-induced  critical points. We
develop a general theory of  such  critical points in 'coherent'
thermodynamics, linking them to the conditions  where the two
 coexisting phases coincide. We show that since the corresponding
critical states  must lie on both the spinodal and the binodal, these two
surfaces in the tensorial space of deformation gradients $\BF$ must be
tangent at the critical points. The latter circumstance allows us to
obtain an explicit analytic characterization of such critical points.
The   general   results obtained in this section can be expected to have important implications for the the design of phase equilibria in highly deformable soft condensed matter,  as well as in   artificial metamaterials, undergoing geometric phase transitions.  Our approach avoids conventional linearization of elastic stresses and strains and is developed  in the  geometrically exact framework of nonlinear elasticity theory.
  
 To illustrate these general results,  we consider,   in Section 4 and Section 5, two  explicit examples showing, that even in the presence of rigidity, depending on a subtle difference in material model,  coherent critical points in the   stress  space may or may not appear. 

The first example presented in  Section 4   is directly related to
swelling phase transition in gels.   We   consider in full detail  an
isotropic solid, exhibiting  the simplest symmetry preserving phase
transition,  with  phases differing only by their
 specific volume.   Such a  transformation in liquids  is known to  exhibit  a classical critical point in the  pressure-temperature space. In this section  we show that an account of   rigidity can give  rise  to a  whole family of non-classical critical points  in the stress space.  More specifically, we  use the Hadamard-Flory model and show that if the elastic shear modulus is sufficiently large, one can   construct    a  loading   protocol which brings  phases with different    reference specific volumes   into each  other continuously while passing around   a rigidity-induced set of  critical points.  We locate this set in both stress and strain spaces, while  retrieving, along the way, a nontrivial  `coherent' generalization of the common tangent construction, which gives rise to nonconvex ground state energy. 

 Our second example, presented in Section 5, reveals the perils of
 geometrical linearization in coherent thermodynamics.  We show that
 if the    geometrically nonlinear  model of an isotropic two-phase
 solid, discussed above,    is replaced by  the more conventional
 geometrically  linearized description,    the   solid-specific
 critical points
 disappear. This example   highlights   the crucial importance of
 using   geometrically exact theories of elasticity in the study of
 coherent phase equilibria in soft solids.  It can be then viewed as a  cautionary
 tale that geometric linearization   can create unphysical artifacts in soft condensed matter.

Finally in Section 6 we summarize our results and present general conclusions.

  \section{ 2. Elastic fluid}

 \begin{figure*}
  \includegraphics[scale=0.17]{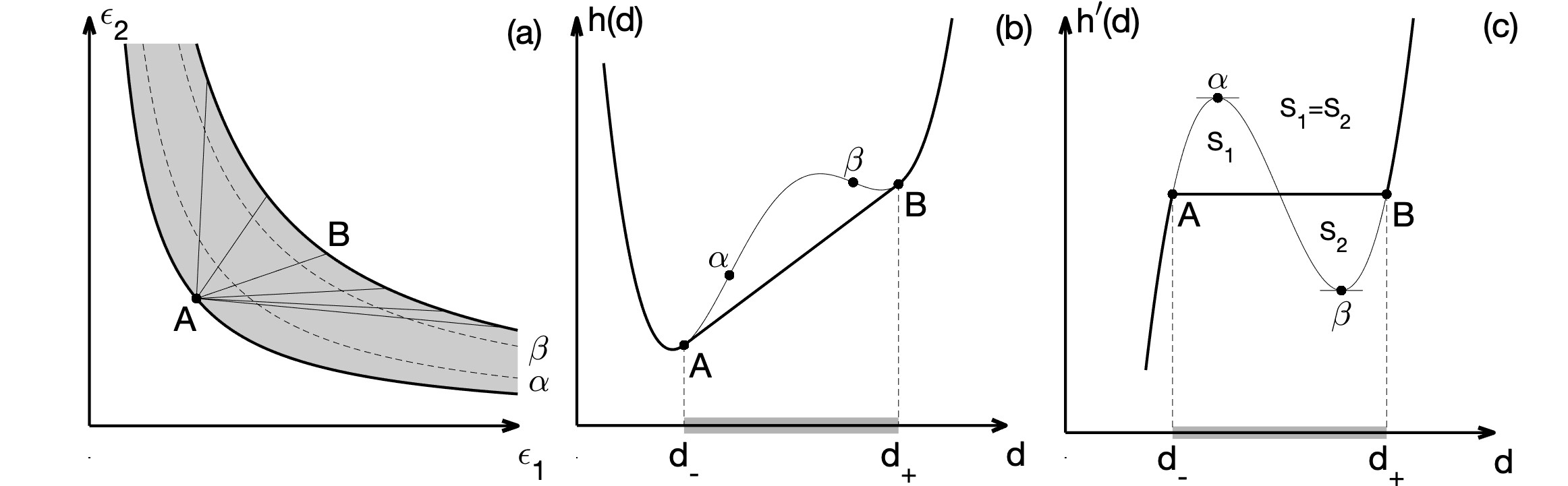} 
 \caption{Phase diagram in the    space of principal strains $ \epsilon_1$ and
   $\epsilon_2$  (a),  double-well energy (b), pressure-volume relation (c)
   for  an elastic liquid.  The relaxed (ground state)  energy and the relaxed
   stress-strain response are shown in (b) and (c) by  thick solid lines.
   Binodal region in (a) is  shaded; dashed lines limit  the spinodal
   region. Here $A$ and $B$ mark  coexisting states   on the binodal, while  $\alpha$ and $\beta$ indicate spinodal points; dashed lines  $\Ga$, $\Gb$ in (a) enclose the spinodal region.}
  \label{fig1}
 \end{figure*}

   To set the stage we first  assume that the  rigidity     can be  neglected  and that the energy density of our `elastic liquid'  is  a function  of the normalized specific volume   $d=\det(\BF)$ only.
   \begin{equation}  \label{ener1}
  W({\BF})=h(d).
  \end{equation}
Since we deal with a volumetric phase transition, the function $h(d)$ is assumed to have   a double well  structure,  see Fig.~\ref{fig1}(b).  This makes the  hydrostatic
  stress (negative of the pressure) $h'(d)$  a non-monotone up-down-up
  function, see Fig.~\ref{fig1}(c);  here and in what follows prime denotes differentiation.

 \subsection{Classical binodal}

  In  such an 'elastic liquid' setting, the
    conditions of phase equilibrium are well known  \cite{lali13v5,daco81}.
The first condition,
\begin{equation}
 \label{eq1}
     \jump{h'(d )}=0,
  \end{equation}  
where  $\jump{A}=A_+-A_-$ denotes the jump, establishes the equality of  pressures in the  two phases. The second, 
  \begin{equation}
 \label{eq2}
     \jump{h(d)}-h'_{\pm}(d)\jump{d}=0,
  \end{equation}    
known as Maxwell equal area construction,  states that the chemical potentials in the two phases must be equal. In the context  of  calculus of variations,  the Gibbs-Maxwell conditions \eqref{eq1}, \eqref{eq2};
the same conditions  naturally reappear in the theory of phase transitions in
elastic bars \cite{erick75}.  
    In accordance with \eqref{eq1},  \eqref{eq2} phase equilibrium
  takes place  at a single (binodal) value of
  pressure.

  While  the above analysis is  fully  three-dimensional, to illustrate the
structure of the  elastic  binodal graphically in the \emph{strain space},
it is convenient to use the 2D version of our model, see Fig.~\ref{fig1}(a).
In view of isotropy and frame indifference, the relevant strain space  is a plane with the
axes representing  singular values of  $\BF$ (principal strains) which we
denote $\epsilon_{1,2}$ in our 2D illustrations.  The domain of phase
coexistence (the binodal region)  in this plane  is shaded   in Fig.~\ref{fig1}(a), where  the hydrostatic (spherical) deformations correspond to the line $\epsilon_{1}= \epsilon_{2}=\epsilon $. Note, that   the elastic liquid   model does not allow us to specify  the  corresponding   two phase microstructures uniquely, and, to stress this point, we indicated, in  Fig.~\ref{fig1}(a),  multiple admissible  connections between a state $A$ and  different states $B$.

   \subsection{Ground state energy}

If we denote  the
  specific  volumes of the two coexisting phases  $d_-\leq d_+$, then
 the  relaxed (ground state) energy density $\tilde W(d)$ is   a convexification of $h(d)$, which  at  $d_{-}\leq d \leq d_{+} $, i.e., where  $h(d)$ differs from its convex hull,  takes the form 
$$\tilde W(d)=
  \frac{d -d_{-}}{d_{+}-d_{-}}h(d_{+})+
\frac{d_{+}-d}{d_{+}-d_{-}}h(d_{-}).$$   
Instead,  at    $d\leq d_-$ and $d\geq d_+$  we have simply   $$\tilde W(d)=h(d).$$
The relaxed energy is shown by a thick solid line in  Fig.~\ref{fig1}(b). The corresponding equilibrium pressure-volume response   is shown in Fig.~\ref{fig1}(c), also by a thick solid line; note that the Maxwell equal area construction is operative in this case.

 \subsection{Classical spinodal}  
 
The two spinodal points  in `elastic liquid'  model, $d_\alpha$ and $d_\beta$,  are defined by the thermodynamic condition \cite{lali13v5},
\begin{equation}
\label{eq5}
h''(d )=0. 
\end{equation}   
They   are located inside the binodal region (between points $A$ and $B$) and are explicitly indicated  in
  Fig.~\ref{fig1}(a,b,c) by letters $ \alpha$ and $ \beta$.
 
  \subsection{Classical critical point}

We adopt, as an operational definition, that at the critical points the
spinodal and the binodal
     meet, and therefore touch; note that  under such an assumption a  critical point is a point on the
spinodal, which is stable. In liquid systems the  critical  points are defined by two equations. 
One of them is \eqref{eq5} since critical points belongs to the spinodal. The second equation, 
indicating the point of tangency of the spinodal and the binodal, is \cite{lali13v5},
  \begin{equation}
\label{eq6}
 h'''(d)=0. 
\end{equation} 
It is easy to see that  the critical state  $d=d_{*}$,  which satisfies  \eqref{eq5}, and \eqref{eq6},  can be also defined  as a configuration  where the distinction between the two coexisting phases, characterized by $d_{+}$
and $d_{-}$ in Fig.~\ref{fig1}(b.c), disappears.

In our  (zero-temperature)  'elastic liquid' model, the two conditions \eqref{eq5},
\eqref{eq6} cannot be satisfied simultaneously for a generic $h(d)$,
and therefore, there are no conventional thermodynamic critical points
in Fig.~\ref{fig1}(a). 
Equations \eqref{eq5}, and \eqref{eq6}  can be both satisfied if we add temperature as a parameter. Then  the inconsistency between \eqref{eq5}, and
  \eqref{eq6} can be overcome given that   the increase of temperature
 is engineered  to move the two wells of the energy density $h(d)$  towards each other.
 Then the value of the temperature,  where the two energy wells 
  coalesce, and therefore  the two equations
  \eqref{eq5}, and \eqref{eq6} have a common solution,  would correspond 
  to   a conventional thermodynamic critical point. We reiterate that our choice of zero temperature is made to ensure that the classical critical points are taken out of the picture, so that we could fully focus on the novel non-classical critical points. 
  
To find  where any distinction between   solid  phases,  coexisting in
  equilibrium,  disappears,  we need to generalize the  equilibrium
  conditions   \eqref{eq1},  \eqref{eq2}, \eqref{eq5}  and \eqref{eq6} for the
  case of general elastic solids. While the analogs of the first three of them
  are basically known, the  fully tensorial counterpart of \eqref{eq6} is not,
  and would have to be derived here.

\section{3. Elastic solid}

In this section we consider an elastic  solid  with nonzero rigidity and    use the methods of  'coherent' thermodynamics to  locate  the coherent binodal and  spinodal in the strain space.  As a bi-product, we also identify the location of  the rigidity-induced critical points.

\subsection{Coherent binodal}

 The 'coherent' analogs of conventional thermodynamic (liquid) equilibrium
 conditions \eqref{eq1} and  \eqref{eq2}  can be  also  viewed as
  equations that the \emph{coexisting} deformation gradients
 $\BF_{\pm}$ must satisfy. Here it is implied that a continuous, piecewise  affine
 deformation $\By_{\pm}(\Bx)$, characterized by  the two gradients $\nabla
 \By_{\pm}(\Bx)=\BF_{\pm}$, which are separated by a plane of discontinuity
 of the deformation gradients, is elastically stable.

 To derive the corresponding  conditions, we start with the reminder that  the relevant  consequence of the elastic stability of the homogeneous configuration
$\By(\Bx)=\BF\Bx$ with respect to all perturbations $\Tld{\By}$ that agree
with $\By(\Bx)$ on the boundary and are uniformly close to $\By(\Bx)$ is the
Weierstrass condition  \cite{mcsh31,graves39,morr52}, 
\begin{equation}
  \label{Weierstrass}
  W(\BF+\Ba\otimes\Bn)\ge W(\BF)+W_{\BF}(\BF)\Bn\cdot\Ba
\end{equation}
for all vectors $\Ba$ and $|\Bn|=1$;  here and in what follows 
the \emph{tensorial} subscripts indicate partial
differentiation with respect to the components of the corresponding
tensorial variables.
 In physical terms, condition \eqref{Weierstrass}
  expresses stability of the homogeneous configuration with respect to
  \emph{nucleation} of coherent  lamina of the new phase
  \cite{ball7677,dufo94}. The `liquid' analog of   \eqref{Weierstrass} is the
  condition of convexity of the function $h(d)$ from \eqref{ener1} at the point $d$.  

Geometrically, inequality \eqref{Weierstrass} partitions the  
space of deformation gradients $\BF$ into two regions: the region $\CB_{\CW}$,
where the Weierstrass necessary condition \eqref{Weierstrass} fails, all of
whose points are definitively unstable, and its complement, where stability
and instability are more difficult to distinguish. We will call the surface
  $\Md\CB_{\CW}$ the \emph{Weierstrass binodal}. Since
  \eqref{Weierstrass} is a necessary condition of elastic stability
  (i.e., of quasiconvexity of  $W(\BF)$), the region $\CB_{\CW}$ lies inside the coherent binodal
  region $\CB$. The latter comprises the set of all elastically unstable homogeneous (affine) 
  configurations.  
  Therefore, all stable points on the Weierstrass
  binodal must necessarily lie on the coherent binodal $\Md\CB$.
  The notion of (coherent)  binodal was   introduced  in \cite{grtrpe,grtrmms}, where it was also explained  why such a  binodal  marks  the boundary of
quasiconvexity of the energy density.

While
the simple constructive characterization of the coherent binodal is hardly possible, 
the Weierstrass binodal is relatively easy to describe
algebraically. Its equations can be found by regarding inequality
\eqref{Weierstrass} as the requirement that the pairs $(\Ba,\Bn)=(0,\Bn)$,
where $\Bn$ is an arbitrary unit vector, are \emph{absolute} minimizers of
\begin{equation}
  \label{jumpset0}
\GD(\Ba,\Bn)=W(\BF+\Ba\otimes\Bn)-W_{\BF}(\BF)\Bn\cdot\Ba.
\end{equation}
Indeed, when $\BF\in\CB_{\CW}$, and this condition fails, there exist $\Ba\not=0$ and $|\Bn|=1$ for which
$\GD(\Ba,\Bn)<W(\BF)$. 

Note next, that on the Weierstrass binodal $\Md\CB_{\CW}$ we expect  the
  existence of $(\Ba,\Bn)$, such that $\GD(\Ba,\Bn)=W(\BF)$,
with $(\Ba,\Bn)$ being a global minimizer of $\GD$. Then, for all $\BF\in\Md\CB_{\CW}$ the following necessary conditions must hold
\begin{eqnarray} \label{3}
&&\GD_{\Ba}(\Ba,\Bn)=0, \\
\label{4}
&& \GD_{\Bn}(\Ba,\Bn)=0,\\
\label{5}
&&\GD(\Ba,\Bn)=W(\BF).
\end{eqnarray}
Introducing notation $\BF_{-}=\BF$, $\BF_{+}=\BF+\Ba\otimes\Bn$,
which ensures that $\BF_{-}$ and $\BF_{+}$ are values of the gradient
of a \emph{continuous} displacement field, we obtain the Hadamard kinematic compatibility condition \cite{ogdenbk97,gfa10,podi13,silh13,krro19},
\begin{equation}
  \label{jumpset1}
    \jump{\BF}=\Ba\otimes\Bn,
\end{equation}
  where  $\Ba$ is the  shear vector, and 
$\Bn $ is the unit   normal to the  phase boundary.  We can then rewrite  equations \eqref{3}--\eqref{5} in the  form of a system
 \begin{eqnarray} \label{31}
&& \jump{\BP}\Bn=0,\\
\label{32}
&&\jump{\BP}^{T}\Ba=0,\\ \label{33} 
&&\jump{W}-\BP_{\pm}\Bn\cdot\Ba\equiv  
\jump{W}-\lump{\BP}\Bn\cdot\Ba=0.
\end{eqnarray}
where $\BP_{\pm}=W_{\BF}(\BF_{\pm})$  and  $\lump{A} =(1/2) (A_++A_-)$.  

The first equation \eqref{31} in this system is the classical traction
continuity condition \cite{podi13,krro19}.  The last equation \eqref{33} is a
generalization of the Maxwell equilibrium condition \cite{eshe70};  the analogs of both of these conditions are already present in the theory
of fluid equilibria, see \eqref{eq1} and \eqref{eq2}. 

Note that the 
 conditions  \eqref{31}, \eqref{33}  can be also viewed as tensorial analogs of the classical Weierstrass-Erdmann
conditions on broken extremals \cite{erdm1877, gesi00,lurie93,cherkbk,alla02}.
They are well known in the calculus of variations and have been also extensively used  in coherent thermodynamics of elastic phase transitions  \cite{robin74,know79,james81,fomcs83,gurtin83,grin86,rosa92,fried93,bhat03,frei07,silh13}.

Instead, condition \eqref{32}  is \emph{solid-specific} and, in this sense, is  rigidity-induced, as it does not exist in classical thermodynamics of liquid phases. Not being relevant in the simplest scalar problem studied by Weierstrass, it has been also overlooked for a long time in the context of calculus of variations.
While the   necessity of additional equalities on smooth broken extremals has been felt  for a long time \cite{Lurie:1970:ODR,roko76,gurtin83,kufr88,sill88,rosl01,silh05,lop07},
in the present general form condition \eqref{32}  was first obtained only recently  \cite{grtrpe}. As it follows from its derivation, see equation \eqref{4},  the condition  \eqref{32}    emerges  from testing the equilibrium phase coexistence   against local re-orientations of the phase boundary viewed as a surface of discontinuity of the deformation gradients, see   \cite{grtrnc} for further  details.

The complete system of equations   \eqref{jumpset1}--\eqref{33}  
describes a  \emph{jump set}, a hyper-surface (surface of codimension
1) in the phase space (space of deformation gradients $\BF$) which
contains  the Weierstrass binodal $\Md\CB_{\CW}$, but may
also contain  branches that  lie inside $\CB_{\CW}$.
In this sense the  Weierstrass binodal  delineates the outer
envelope of the \emph{jump set}, stable parts of which belong to the
coherent binodal. By emphasizing this point we stress that, while it is never  an issue in liquid  systems,  some of the
ensuing thermodynamic  phase equilibria can be still elastically
unstable.

One way to distinguish stable 
points on $\Md\CB_{\CW}$ from those inside $\CB_{\CW}$ is to check the non-negativity of the Hessian
\[
\BH=\mat{\GD_{\Ba\Ba}}{\GD_{\Ba\Bn}}{\GD_{\Bn\Ba}}{\GD_{\Bn\Bn}}.
\]
It turns out that on the part of the hypersurface  \eqref{jumpset1}--\eqref{33},   that
satisfies $\BH>0$, each
$\BF=\BF_{-}$ has a uniquely defined $\BF_{+}$ that depends smoothly
on $\BF_{-}$.  Note that  the positive definiteness of $\BH$ has to be understood, while accounting  for  its geometrical degeneracy induced by the fact that  Weierstrass condition depends on $\Ba$ and $\Bn$  only through the  combination $\Ba\otimes \Bn$.

In what follows we denote by $\CJ$ the part of the hypersurface
\eqref{jumpset1}--\eqref{33} that satisfies $\BH>0$.  In our example of
Hadamard-Flory solid, discussed in Section~4, the whole
hypersurface defined by \eqref{jumpset1}--\eqref{33} will have this
property. It will also coincide with the coherent binodal. In our
geometrically linear example considered in Section~5 the
surface defined by \eqref{jumpset1}--\eqref{33} also coincides with the
coherent binodal, but has $\BH=0$ at each point. As we are going to see, the
degeneracy is, in part, due to the enhanced compatibility of the linearized strain.

\subsection{Coherent spinodal}
 One of the elementary consequences of the Weierstrass stability condition
\eqref{Weierstrass} is obtained by restricting $\Ba$ to a small \nbh\ of
zero. From the expansion
\begin{eqnarray*} 
 &&W(\BF+\Ba\otimes\Bn)-W(\BF)-W_{\BF}(\BF)\Bn\cdot\Ba=\\
&&\hf\BA(\BF;\Bn)\Ba\cdot\Ba+O(|\Ba|^{3}),
\end{eqnarray*} 
we obtain a corollary of \eqref{Weierstrass},  known as the Legendre-Hadamard
condition, or the ellipticity condition for equations of elastostatics, e.g. \cite{silh13,krro19,Dak08}, 
\begin{equation}
  \label{LH}
 \BA(\BF;\Bn)\Ba\cdot\Ba\ge 0,
\end{equation}
for all $|\Ba|=1$ and $|\Bn|=1$. Here
$\BA(\BF;\Bn)$ is the acoustic tensor  defined by its quadratic form 
$$\BA(\BF;\Bn)\Ba\cdot\Ba=W_{\BF\BF}(\BF)[\Ba\otimes\Bn,\Ba\otimes\Bn],$$
or  in index notation, by the formula
$$A_{ij}(\BF;\Bn)=W_{F^i_{\Ga}F^j_{\Gb}}(\BF)n_{\Ga}n_{\Gb},$$
 where the summation over repeated indexes is assumed. 
 
 Just as in the case of the  Weierstrass binodal, it is advantageous to
view \eqref{LH} geometrically as a partitioning of the phase space into two
regions:  $\CS$, where \eqref{LH} fails (coherent spinodal region), and its
complement. The boundary $\Md\CS$  can be then  interpreted 
as the coherent spinodal, see  \cite{grtrmms} for further details.
 
 According to  \eqref{LH} all the
eigenvalues of real symmetric matrices $\BA(\BF;\Bn)$ have to be nonnegative for all $|\Bn|=1$.
 While the computation of eigenvalues of real symmetric
 matrices is standard, the verification of their nonnegativity for an infinite
 family of unit vectors $\Bn$ can make the task
 challenging. 
 
To overcome this difficulty, we first emphasize that when
$\BF\in\Md\CS$, there exists a unit vector $\Bn_{0}$, such that at least
one of the eigenvalues of the acoustic tensor $\BA(\BF;\Bn_{0})$ becomes zero,
while others remain positive. In addition, the eigenvalues of $\BA(\BF;\Bn)$
must be nonnegative  for \emph{all} unit vectors $\Bn$. Denoting by $\Ba_{0}$
the unit eigenvector of $\BA(\BF;\Bn_{0})$ corresponding to the zero eigenvalue, we
 obtain our first set of equations for the coherent spinodal
  \begin{equation}
    \label{spineq1}
\BA(\BF;\Bn_{0})\Ba_{0}=0.    
  \end{equation}
The second set is obtained from the observation that
$\GL(\Bn)=\BA(\BF;\Bn)\Ba_{0}\cdot\Ba_{0}$ achieves its global minimum of 0 at
$\Bn=\Bn_{0}$, so that
\begin{equation}
  \label{spineq2}
 \BA^*(\BF;\Ba_{0})\Bn_{0}=0.  
\end{equation}
where $ \BA^*(\BF;\Ba)$ is the co-acoustic tensor  defined by its
quadratic form
$$\BA^*(\BF;\Ba)\Bn\cdot\Bn=W_{\BF\BF}(\BF)[\Ba\otimes\Bn,\Ba\otimes\Bn]$$
or in index notation by the formula
$$A^{*\Ga\Gb}(\BF;\Ba)=W_{F^i_{\Ga}F^j_{\Gb}}(\BF)a^{i}a^{j},$$ 
where we again assumed the summation over repeated indexes.

Equations  \eqref{spineq1}--\eqref{spineq2}, satisfied by all
points on the coherent spinodal, can be viewed as the  tensorial  analogs  of \eqref{eq5}. Upon  elimination of the unit vectors $\Bn_{0}$ and $\Ba_{0}$, they  describe a hypersurface in the phase
space of deformation gradients $\BF$. Indeed, 
in $n$ space dimensions there are $n^{2}+(n-1)+(n-1)$
unknowns in the set of tensorial variables
$(\BF,\Ba_{0},\Bn_{0})$, since both $\Ba_{0}$ and $\Bn_{0}$ are unit vectors. Equation  \eqref{spineq1} has $n$ scalar restrictions, as does equation  \eqref{spineq2}. 
However, there is one
scalar relation between the two equations, since
$$
 \BA^*(\BF;\Ba_{0})\Bn_{0}\cdot\Bn_{0}=\BA(\BF;\Bn_{0})\Ba_{0}\cdot\Ba_{0}.
$$
Thus, the space of solutions $(\BF,\Ba_{0},\Bn_{0})$ of
\eqref{spineq1}--\eqref{spineq2} is
$n^{2}-1$ dimensional. Therefore, under some basic nondegeneracy assumptions, we can claim
that the solution set has
$n^{2}-1$ dimensional projection onto the phase space of
deformation gradients $\BF$. Note, that while all points on the
coherent spinodal solve the system 
  \eqref{spineq1}--\eqref{spineq2},  such a  system by itself may also have other solution
  branches, all in the interior of $\CS$.

\subsection{ Coherent critical points}
 Recall that we defined \emph{coherent critical points} as the
points of intersection, and therefore tangency, of the coherent binodal and the
coherent spinodal. This operational definition implies that critical points are exactly the stable points on the
spinodal, in particular, the corresponding states (deformation gradients $\BF_{*}$)   must satisfy the necessary
condition of Weierstrass \eqref{Weierstrass}.

  Using $\Ba=t\Ba_{0}$ and $\Bn=\Bn_{0}$ in the
  Weierstrass condition \eqref{Weierstrass} we obtain
\begin{eqnarray*} 
 &&W(\BF_*+t\Ba_{0}\otimes\Bn_{0})-W(\BF_*)-tW_{\BF}(\BF)\Bn_{0}\cdot\Ba_{0}=\\
&&\frac{t^{3}}{6}W_{\BF\BF\BF}(\BF_*)[\Ba_{0}\otimes\Bn_{0},\Ba_{0}\otimes\Bn_{0},\Ba_{0}\otimes\Bn_{0}]
+O(t^{4}),
\end{eqnarray*}
where we used that  
\begin{equation}
  \label{acou0}
\BA(\BF_*;\Bn_{0})\Ba_{0}\cdot\Ba_{0}=0.
\end{equation}
Furthermore, given that the leading term above has an odd power of $t$, inequality
\eqref{Weierstrass} implies that we must have
\begin{equation}
 \label{cpe} W_{\BF\BF\BF}(\BF_*)[\Ba_{0} \otimes\Bn_{0},\Ba_{0} \otimes\Bn_{0},\Ba_{0}
 \otimes\Bn_{0}]=0.
  \end{equation}
This equation shows that if $\BF_*\in\Md\CS$  is stable, then 
$\BF$ must satisfy
\eqref{spineq1}, \eqref{spineq2}, and \eqref{cpe}.  Upon
elimination of unit vectors $\Ba_{0}$ and $\Bn_{0}$, we are left 
with two scalar equations, locating the state $\BF_*$ on a co-dimension one subvariety
of the spinodal. 

 Let us now demonstrate that critical points admit a more
  classical description as points in the phase space, where the distinction
  between the two coexisting phases disappears.  
  Recall that  if $\BF_{*}$ is a point of tangency between the coherent binodal and
  the coherent spinodal, then, it must also be a point of tangency
  between the Weierstrass binodal $\Md\CB_{\CW}$ and the coherent spinodal. Indeed,
  on the one hand, all points in the complement of the coherent
  binodal region $\CB$ are stable, and therefore satisfy
the Weierstrass necessary condition \eqref{Weierstrass}. On the hand,
all points in the coherent spinodal region $\CS$ fail
\eqref{Weierstrass}. Therefore, the Weierstrass binodal must pass
between $\Md\CB$ and $\Md\CS$, separating $\CS$ from the complement of
$\CB$. Hence, $\Md\CB_{\CW}$ must be tangent to
$\Md\CS$ at the point $\BF_{*}$, where $\Md\CB$ and $\Md\CS$ touch. 
  
Suppose $\BF_{-}\in\Md\CB_{\CW}$ is an arbitrary sequence or family of points
on $\Md\CB_{\CW}$, such that $\BF_{-}\to \BF_{*}$. We claim that, generically,
the corresponding $\BF_{+}$ on $\Md\CB_{\CW}$, solving the jump set equations
\eqref{jumpset1}--\eqref{33} must also satisfy $\BF_{+}\to\BF_{*}$. This is
based on the observation that generically the failure of the pairs $(0,\Bn)$
to be global minimizers of $\GD(\Ba,\Bn)$ occurs either by $(0,\Bn)$ failing
to be a local minimizer or by the appearance of an additional global minimizer
$(\Ba,\Bn)$, \emph{but not both at the same time.} In this case, if $\BF_{+}$
does not converge to $\BF_{*}$, when $\BF_{-}$ does, then
$\BF_{+}\to\BF_{*}+\Ba\otimes\Bn$, for some $\Ba\not=0$ and $|\Bn|=1$. But
then, $(\Ba,\Bn)$ is a global minimizer of $\GD(\Ba,\Bn)$, when
$\BF=\BF_{*}$. It follows that as $\BF_{*}$ moves into the spinodal region,
the minimality of $\{(0,\Bn):|\Bn|=1\}$ is violated both locally and globally,
which is a non-generic behavior. We conclude that generically, if
$\BF_{-}\to\BF_{*}$, then $\BF_{+}\to\BF_{*}$, as well.  Thus, each critical
point must also be the common limit of a family of coexisting states
$\BF_{+}$, $\BF_{-}$, satisfying phase equilibrium equations
\eqref{jumpset1}--\eqref{33}.

Let us show that conversely, any point $\BF_{*}$ on the Weierstrass binodal
(the jump set) at which the distinction between $\BF_{+}$ and $\BF_{-}$
disappears has to satisfy equations \eqref{spineq1}, \eqref{spineq2},
\eqref{cpe}. Specifically, we assume that for an arbitrary smooth curve
$\BF_{-}(t)$ on the Weierstrass binodal, such that $\BF_{-}(0)=\BF_{*}$,
there exists a corresponding smooth curve $\BF_{+}(t)$, such that
$\BF_{+}(0)=\BF_{*}$, and $\BF_{\pm}(t)$ solve the jump set equations
\eqref{jumpset1}--\eqref{33} for every (sufficiently small) $t$. While all
points $\BF_{\pm}(t)$ are constrained by \eqref{jumpset1}--\eqref{33} when
$t\not=0$, at the critical point the system \eqref{jumpset1}--\eqref{33}
trivializes, as it is automatically satisfied whenever $\BF_{+}=\BF_{-}$. To
overcome such an obstacle we need to ``zoom in'' to the \nbh\ of the critical
point $\BF_{*}$ and compute the leading order expansions of equations
\eqref{jumpset1}--\eqref{33} around it.

First of all, it is important to ensure that the
curve $\BF(t)$ has a non-singular parametrization. The most convenient choice
of the parameter would be $$s=|\BF_{+}(t)-\BF_{-}(t)|,$$
where $|\BF|$ denotes the Frobenius norm of a matrix $\BF$, and we use the parameter $s$ in what follows. In this case we can
write $\jump{\BF}=s\hat{\Ba}(s)\otimes\Bn(s)$, where $|\hat{\Ba}(s)|=1$, and $|\Ba|$
denotes the usual Euclidean length of a vector $\Ba$.
Differentiating (in $s$) equation \eqref{jumpset1}, we obtain at $s=0$,
$$\jump{\dot{\BF}}=\hat{\Ba}(0)\otimes\Bn(0).$$ 
Differentiating equations \eqref{31}, \eqref{32} we obtain, at $s=0$,
equations \eqref{spineq1}, \eqref{spineq2}, respectively, with
$\Ba_{0}=\hat{\Ba}(0)$ and $\Bn_{0}=\Bn(0)$.   Thus, we have shown that any
point $\BF_{*}\in\Md\CB_{\CW}$, at which the distinction between phases disappears must
solve the system of equations
\eqref{spineq1}--\eqref{spineq2}. In particular, $\BF_{*}$ must lie in the closure of
$\CS$. Since we have seen that the Legendre-Hadamard condition \eqref{LH} is
a consequence of the Weierstrass condition \eqref{Weierstrass}, spinodal must
necessarily lie in the closure of $\CB$. We conclude that
$\BF_{*}\in\Md\CB_{\CW}$ must also lie on the spinodal
$\Md\CS$.

Let us show now that the leading term in the expansion of the
remaining equation \eqref{33} coincides with \eqref{cpe}. In order to make the calculations more compact we introduce the ``directional derivatives'' notation 
for multiple derivatives with respect to $\BF$:
\begin{eqnarray*}
&&W_{F_{\Ga}^{i}}(\BF)H^{i}_{\Ga}=P[\BH],\\
&& W_{F_{\Ga}^{i}F_{\Gb}^{j}}(\BF)H^{i}_{\Ga}G^{j}_{\Gb}=L[\BG,\BH],\\
&&W_{F_{\Ga}^{i}F_{\Gb}^{j}F_{\Gg}^{k}}(\BF)H^{i}_{\Ga}G^{j}_{\Gb}K_{\Gg}^{k}=M[\BK,\BG,\BH],
 \end{eqnarray*}
  where we again assumed the summation over repeated indexes.
The first derivative of the Maxwell condition \eqref{33} in $s$
would then be
$$
\jump{P[\dot{\BF}]}-L_{\pm}[\dot{\BF}_{\pm},\jump{\BF}]-P_{\pm}[\jump{\dot{\BF}}]=0.
$$
 Since $\BF_{\pm}(0)=\BF_{*}$ the \lhs\ of the equation above is zero
 at $s=0$. Hence, we
must take another derivative to find the leading term. Differentiating the
\lhs\ above, we obtain  
\begin{eqnarray*}
&&  \jump{L[\dot{\BF},\dot{\BF}]}+\jump{P[\ddot{\BF}]}-
M_{\pm}[\dot{\BF}_{\pm},\dot{\BF}_{\pm},\jump{\BF}]-\\
&&L_{\pm}[\ddot{\BF}_{\pm},\jump{\BF}]-2L_{\pm}[\dot{\BF}_{\pm},\jump{\dot{\BF}}]-
P_{\pm}[\jump{\ddot{\BF}}]=0.
\end{eqnarray*}
 Now, at $s=0$ we obtain
  $L_{*}[\jump{\dot{\BF}},\jump{\dot{\BF}}]=0$, where $L_{*}$ denotes $W_{\BF\BF}(\BF_{*})$.
As we have already shown,
\begin{equation}
  \label{jF}
    \jump{\dot{\BF}}=\Ba_{0}\otimes\Bn_{0},
\end{equation}
and since $\BF_{*}$, $\Ba_{0}$, $\Bn_{0}$ satisfy \eqref{spineq1}, then
\[
L_{*}[\jump{\dot{\BF}},\jump{\dot{\BF}}]=\BA(\BF_{*};\Bn_{0})\Ba_{0}\cdot\Ba_{0}=0.
\]
Hence, another differentiation in $s$ at $s=0$ needs to be computed. This
time, we will not write the expression for the third derivative 
of \eqref{33}, but only the value
at $s=0$. After performing obvious cancellations, we obtain
\begin{eqnarray*}
&&\jump{M_{*}[\dot{\BF},\dot{\BF},\dot{\BF}]}+3\jump{L_{*}[\ddot{\BF},\dot{\BF}]}
-3M_{*}[\dot{\BF}_{\pm},\dot{\BF}_{\pm},\jump{\dot{\BF}}]\\
&&-3L_{*}[\ddot{\BF}_{\pm},\jump{\dot{\BF}}]
-3L_{*}[\dot{\BF}_{\pm},\jump{\ddot{\BF}}]=0.
\end{eqnarray*}
Here  $M_{*}$ denotes $W_{\BF\BF\BF}(\BF_{*})$.  
Expanding the jump notation it is easy to check that
\begin{multline*}
\jump{L_{*}[\ddot{\BF},\dot{\BF}]}-L_{*}[\ddot{\BF}_{\pm},\jump{\dot{\BF}}]
-L_{*}[\dot{\BF}_{\pm},\jump{\ddot{\BF}}]=\\
\mp L_{*}[\jump{\ddot{\BF}},\jump{\dot{\BF}}],
\end{multline*}
and similarly
\begin{multline*}
\jump{M_{*}[\dot{\BF},\dot{\BF},\dot{\BF}]}-
3M_{*}[\dot{\BF}_{\pm},\dot{\BF}_{\pm},\jump{\dot{\BF}}]=\\
\mp3M_{*}[\dot{\BF}_{\pm},\jump{\dot{\BF}},\jump{\dot{\BF}}]+
M_{*}[\jump{\dot{\BF}},\jump{\dot{\BF}},\jump{\dot{\BF}}].
\end{multline*}
The third derivative of the Maxwell relation then simplifies to
\begin{multline*}
\mp3M_{*}[\dot{\BF}_{\pm},\jump{\dot{\BF}},\jump{\dot{\BF}}]+
M_{*}[\jump{\dot{\BF}},\jump{\dot{\BF}},\jump{\dot{\BF}}]\mp\\
3L_{*}[\jump{\ddot{\BF}},\jump{\dot{\BF}}]=0
\end{multline*}
Taking into account that
\begin{eqnarray*}
  \label{jumps1}
  &&\jump{\dot{\BF}}=\Ba_{0}\otimes\Bn_{0},\\
  \label{jumps2}
  &&\jump{\ddot{\BF}}=2(\hat{\Ba}'(0)\otimes\Bn_{0}+\Ba_{0}\otimes\Bn'(0)),
\end{eqnarray*}
where $\hat{\Ba}'(0)$ and $\Bn'(0)$ denote derivatives of $\hat{\Ba}(s)$ and
$\Bn(s)$ at $s=0$, we compute
\begin{multline*}
L_{*}[\jump{\ddot{\BF}},\jump{\dot{\BF}}]=2\BA(\BF_{*},\Bn_{0})\Ba_{0}\cdot\hat{\Ba}'(0)+\\
2\BA^{*}(\BF_{*},\Ba_{0})\Bn_{0}\cdot\Bn'(0)=0,
\end{multline*}
where we used  \eqref{spineq1} and \eqref{spineq2}. We conclude that the leading term in
the Maxwell relation is
\begin{equation}
  \label{limax}
  M_{*}[\jump{\dot{\BF}},\jump{\dot{\BF}},\jump{\dot{\BF}}]\mp
3M_{*}[\dot{\BF}_{\pm},\jump{\dot{\BF}},\jump{\dot{\BF}}]=0.
\end{equation}
Now adding both equations (one for each sign) we obtain
\[
M_{*}[\jump{\dot{\BF}},\jump{\dot{\BF}},\jump{\dot{\BF}}]=0,
\]
which, due to \eqref{jF}, coincides with \eqref{cpe},  showing that the two
interpretations of criticality, (i) as the points of tangency of the coherent binodal and
the coherent spinodal, and (ii) as points, where the distinction between coexisting phases disappear, coincide.

 Observe now, that \eqref{cpe} is not the only consequence of
  \eqref{limax}. We also obtain
\[
M_{*}[\dot{\BF}_{\pm},\jump{\dot{\BF}},\jump{\dot{\BF}}]=0.
\]
Since $\BF_{-}(s)$ was an arbitrary curve on $\Md\CB_{\CW}$ passing through $\BF_{*}$ at $s=0$,
the normal to the Weierstrass binodal at $\BF_{*}$ can be identified with the functional
\begin{equation}
  \label{NWB}
  \BN_{*}=M_{*}[\Ba_{0}\otimes\Bn_{0},\Ba_{0}\otimes\Bn_{0},_{-}]
\end{equation}
in the sense that 
\[
M_{*}[\Ba_{0}\otimes\Bn_{0},\Ba_{0}\otimes\Bn_{0},\BT]=0
\]
for any tangent $\BT$ to the Weierstrass binodal at $\BF_{*}$.

Next, we show that equation \eqref{cpe} is the only equation, in addition to
\eqref{spineq1}, \eqref{spineq2}, that all critical points must satisfy. At
the first glance there should be many more equations, since when two generic
hypersurfaces in an $n$-dimensional space with equations $F(\Bx)=0$ and
$G(\Bx)=0$ touch, the equations for the point of tangency is not only the two
equations above, but also $\Grad F=\Gl\Grad G$, expressing the collinearity of
the normals to the two surfaces at the point of tangency. This gives an a
priori overdetermined (since tangency of two surfaces is not a generic
configuration) system of $n+2$ equations on $n+1$ unknowns $\Bx$, $\Gl$. This
overdeterminacy is the mathematical underpinning of our claim that that when
$\BF_{-}\to\BF_{*}\in\Md\CB_{\CW}\cap\Md\CS$, then, generically,
$\BF_{+}\to\BF_{*}$. 
 
Therefore, in the case of the tangency of the binodal and the spinodal, we
need to examine the relation between their normals at $\BF_{*}$, in addition
to the equations that place $\BF_{*}$ on both $\Md\CS$ and $\Md\CW_{\CW}$. The
normal to the binodal is given by \eqref{NWB}. Let us compute the normal to
the spinodal at an arbitrary point $\BF_{0}\in\Md\CS$.  To this end we
consider an arbitrary smooth curve $\BF(t)$ on the spinodal, such that
$\BF(0)=\BF_{0}$. For each $t$ there are corresponding unit vectors $\Ba(t)$
and $\Bn(t)$, such that equations \eqref{spineq1}, \eqref{spineq2} are
satisfied by $\BF(t)$, $\Ba(t)$, $\Bn(t)$. Assuming that the functions above
are smooth, and differentiating these equations with respect to $t$ at $t=0$
we will obtain a single scalar linear constraint that the tangent
$\dot{\BF}(0)$ to the spinodal have to satisfy, thereby revealing the normal
to the spinodal. In order to perform the differentiation it will be convenient
to rewrite \eqref{spineq1}, \eqref{spineq2} in a more generic form
\begin{equation}
  \label{spineq1t}
  L[\Ba(t)\otimes\Bn(t),\Ba(t)\otimes\Bu]=0,
\end{equation}
\begin{equation}
  \label{spineq2t}
 L[\Ba(t)\otimes\Bn(t),\Bv\otimes\Bn(t)]=0,
\end{equation}
valid for any choice of vectors $\Bu$ and $\Bv$. 
Differentiating the two equations with respect to $t$ at $t=0$ we obtain
\begin{multline*}
  M[\dot{\BF},\Ba_{0}\otimes\Bn_{0},\Ba_{0}\otimes\Bu]+
L[\dot{\Ba}\otimes\Bn_{0},\Ba_{0}\otimes\Bu]+\\
L[\Ba_{0}\otimes\dot{\Bn},\Ba_{0}\otimes\Bu]+
L(\BF_{0})[\Ba_{0}\otimes\Bn_{0},\dot{\Ba}\otimes\Bu]=0,
\end{multline*}
and
\begin{multline*}
  M[\dot{\BF},\Ba_{0}\otimes\Bn_{0},\Bv\otimes\Bn_{0}]+
L[\dot{\Ba}\otimes\Bn_{0},\Bv\otimes\Bn_{0}]+\\
L[\Ba_{0}\otimes\dot{\Bn},\Bv\otimes\Bn_{0}]+
L[\Ba_{0}\otimes\Bn_{0},\Bv\otimes\dot{\Bn}]=0,
\end{multline*}
respectively. Finally, substituting $\Bu=\Bn_{0}$ or $\Bv=\Ba_{0}$, and taking
equations \eqref{spineq1t}, \eqref{spineq2t} into account we obtain the  relation
$M[\dot{\BF},\Ba_{0}\otimes\Bn_{0},\Ba_{0}\otimes\Bn_{0}]=0$.
Therefore the normal to a generic point on the  spinodal surface 
can be identified with the functional
\begin{equation}
  \label{spinorm}
\BN_{\Md\CS}(\BF_0)= M[\Ba_{0}\otimes\Bn_{0},\Ba_{0}\otimes\Bn_{0},_{-}],
\end{equation}
meaning that
\[
M[\Ba_{0}\otimes\Bn_{0},\Ba_{0}\otimes\Bn_{0},\BT]=0
\]
for any tangent $\BT$ to the spinodal. Comparing \eqref{NWB} and \eqref{spinorm}, we conclude that the normal to $\Md\CB_{\CW}$ at the
  coherent critical point $\BF_{*}$
  is always parallel to the normal to the spinodal at the same point. As a
  consequence, the only equations satisfied by a critical point are
  \eqref{spineq1}, \eqref{spineq2}, and \eqref{cpe} and therefore, the
two surfaces must touch, not at a single point, but along a codimension 2 surface in
the phase space.
 
We conclude that in coherent thermodynamics we should talk about a \emph{critical set}, rather than
a single critical point. The critical
set is the set of stable solutions of the system \eqref{spineq1},
\eqref{spineq2}, \eqref{cpe}, 
which, generically, should be a subvariety of both the binodal and the spinodal
of codimension one in each. More specifically, in 3D  the  coherent critical set,
if it exists, is  a  set of dimension 7 in the 9-dimensional phase space, and in
2D it is  a  set of
dimension 2 in the 4-dimensional phase space. Since for isotropic solids the phase diagram can be drawn in the space of singular values of $\BF$, it will exhibit  the critical
set as  a point in 2D and as a line in 3D. Despite the relatively  high dimensionality of the critical set (generically, it is not a point),  its  dimensional deficiency allows one  to   pass around it  and  reach  continuously from one coherent phase to another. 
  
 \begin{figure*} 
\includegraphics[scale=0.15]{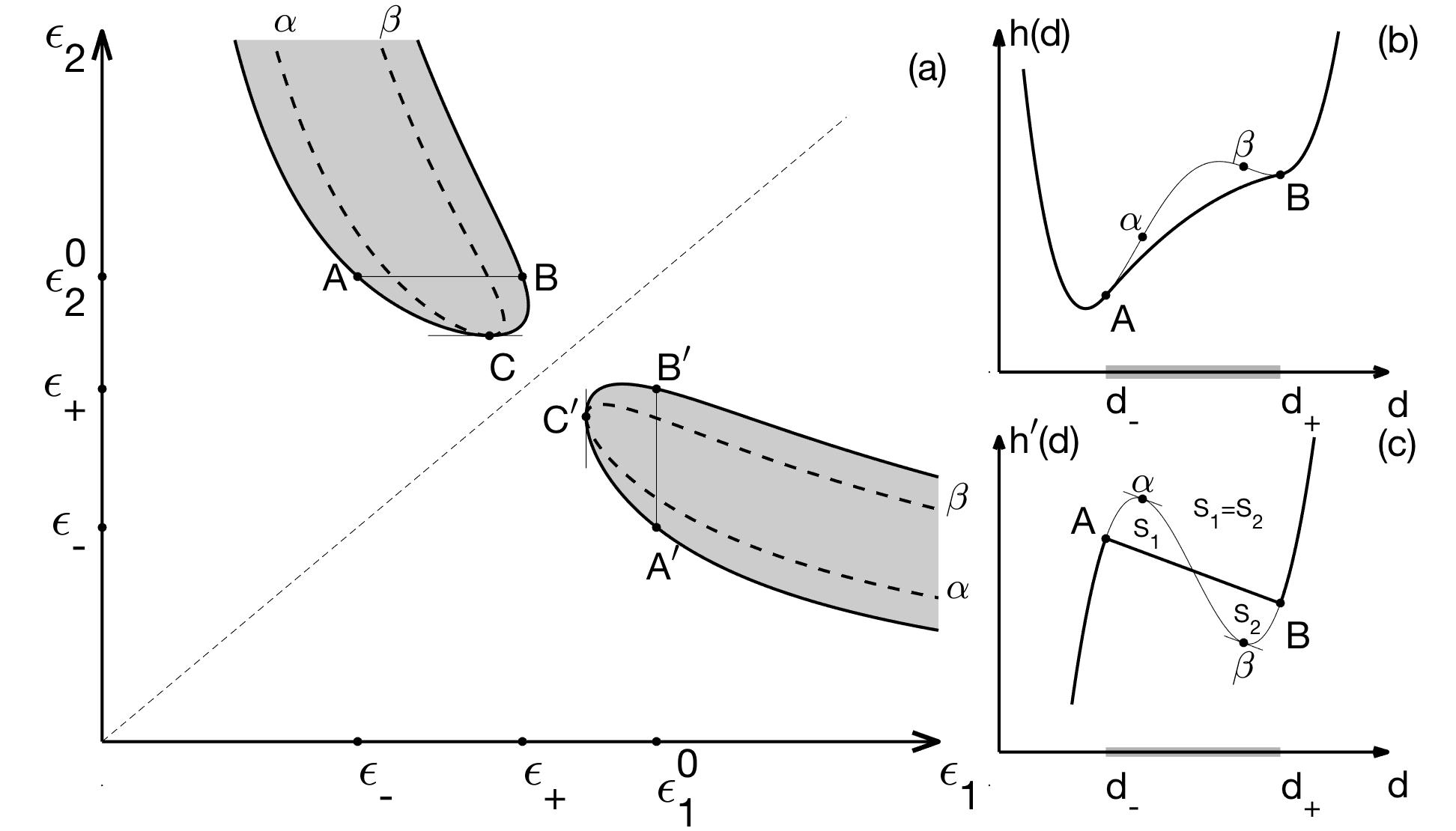}  
 \caption{  (a) Phase diagram in the space of principal strains   for an Hadamard-Flory solid with sufficiently large rigidity. (b) 
Double-well part of the energy
along the line $AB$: $\epsilon_1=d/\epsilon_{2}^{0}$, $\epsilon_{2}=\epsilon_{2}^{0}$.  (c)
The corresponding pressure-volume relation  along the line $AB$. The  double-well
  part $\tilde{W}(\Ge_{1},\Ge_{2})-(\mu/2)(\Ge_{1}^{2}+\Ge_{2}^{2})$ of the relaxed (ground state)  energy and the relaxed
stress-strain response are shown by solid lines.  Binodal region in (a) is
shaded. $AB$ and $A'B'$ are typical pairs of coexisting states, $C$ and $C'$
are critical points. Dashed lines  $\Ga$, $\Gb$ enclose the spinodal region. }
  \label{fig5}
\end{figure*} 

For the elementary examples of coherent critical points, see
\cite{ftz03,butr93}. In what follows we discuss two non-elementary
applications of the obtained formulas, showing that in superficially similar
models of elastic solids coherent critical sets may exist but may also be
prohibited. The latter situation arises when either the system of equations
\eqref{spineq1}, \eqref{spineq2}, \eqref{cpe} has no solutions, or when all of
their solutions are unstable.
 
\section{4. Geometrically nonlinear Hadamard-Flory solid}
To illustrate the   conditions of coherent criticality \eqref{spineq1}, \eqref{spineq2}, \eqref{cpe}, we now consider the simplest generalization of the  'elastic liquid' model which
  accounts for rigidity.  Known as the  Hadamard-Flory model, it is characterized by the energy density \cite{hada03,flor53,onuk88b,ssk90}  
\begin{equation} \label{tr1}
 W(\BF)= h(\det\BF)+(\mu/2)|\BF|^{2},
\end{equation}  
where the first term is  the same double well potential as in the 'elastic liquid' model. The magnitude of the second (shear related)  term in \eqref{tr1} is controlled by the   rigidity   modulus $\mu$;   in gels,  its value  is mostly affected   by  the degree of cross-linking.

 \subsection{Coherent  binodal} 

We take advantage of a known fact \cite{grtrsolid} that to construct the  surface  of phase coexistence (the coherent binodal) for the Hadamard-Flory solid at sufficiently large values of $\mu$, it is sufficient to   consider
simple laminates which are layered mixtures of  two   deformation gradients
$\BF_{\pm}$.  The full set of equations such a pair needs to
  satisfy is given by \eqref{jumpset1}--\eqref{33}.
   These equations define the jump set for the Hadamard-Flory solid
  \eqref{tr1}. 
Observe first, that  
kinematic compatibility condition \eqref{jumpset1} implies
   $$d_{+}=d_{-}(1 +   {\BF} _{-}^{-T}{\Bn} \cdot{\Ba}),$$ 
where again $d_{\pm}=\det\BF_{\pm}$. 
Using this equality, equation \eqref{jumpset1}, and the  formula for the Piola stress tensor 
$$
 \BP =W_{\BF}(\BF)=\mu\BF+d h'(d) \BF^{-T},
$$
 we can reduce the condition of  traction continuity  \eqref{31} to 
$$
  \Ba=-(\jump{h'}d_-/\mu)  \BF_{-}^{-T}\Bn.
$$
Then, using  \eqref{32} 
we find  that $\Bn$ must be an eigenvector of the Cauchy-Green strain tensor  
$$\BC_{-}=\BF_{-}^{T}\BF_{-},$$  and also that $\BC_{+}$ and $\BC_{-}$ are related by
$$\BC_{+}=\BC_{-}+\left(|\Ba|^{2}-(2\jump{h'}d_{-}/\mu)\right) \Bn\otimes
\Bn.$$ This implies that $\BC_{+}$ and $\BC_{-}$ are
simultaneously diagonalizable. Therefore, there  exist coordinate frames in
both Lagrangian and Eulerian spaces in which
both deformation gradients $\BF_{\pm}$ are simultaneously diagonal, and differ by a single
eigenvalue $\Ge_{\pm}$, corresponding to the eigenvector $\Bn$. 

If we now denote
the product of the remaining common eigenvalues of $\BF_{\pm}$  (i.e. common singular
values, in an arbitrary frame) by $\Ge_{0}$, then
we can characterize the set of coexisting deformation gradients  by the
equation 
\begin{equation}
  \label{js}
 \epsilon_0^2 \jump{h'}+\mu\jump{d}=0,  
\end{equation}
 where the relation between the values
 $d_{\pm}$ are  found from \eqref{33},
 which now has the form
 \begin{equation}
   \label{maxwell}
  \jump{h}-\lump{ h'} \jump{d}=0.   
 \end{equation}
 Note that equations \eqref{js}, \eqref{maxwell}
  can  be  interpreted geometrically as equality of areas between the  (tilted)  line with the slope
\begin{equation}
  \label{slopex}
  \jump{h'}/\jump{d}=-\mu/\epsilon_0^2
\end{equation}
 and the graph of $h'(d)$, see Fig.~\ref{fig5}(c). The nonzero slope in
 \eqref{slopex} implies that the classic (not `tilted') Maxwell common tangent construction no longer
 applies, and the relaxed energy, shown in Fig.~\ref{fig5}(b), is no longer convex, due to the extensive mixing effects of purely elastic origin. 

The foregoing analysis shows that we can draw a phase diagram in the
two-dimensional plane, where one coordinate denotes one of the singular values of
the deformation gradient $\BF$, while the other, the product of the remaining
ones. However, in 2D, both coordinates in such a phase diagram have the
meaning of singular values of $\BF$, making the interpretation of the figures
more natural. Hence, one can view our figures as 2D phase diagrams in the
space of singular values of $\BF$, while keeping in mind that they can also be
interpreted as phase diagrams in arbitrary number of space dimensions.
In the regime of
sufficiently large $\mu$ the set of coexisting states takes the form of
two separate curves in the
  $(\Ge_{1},\Ge_{2})$-space, shown in Fig.~\ref{fig5}(a).  The presence of two
symmetric subdomains  $ \epsilon_{1}>\epsilon_{2} $  and $\epsilon_{1}<\epsilon_{2}$ reflects  the symmetry of the system  with respect
to the interchange of singular values of $\BF$.  In three space
dimensions, the phase diagram would be represented by a single, say upper,
curve, while the horizontal and vertical axes would be labelled as $\Ge$, and
$\Ge_{0}$, denoting one of the singular values of $\BF$, and
the product of the remaining ones, respectively. In what follows our figures
will always be drawn in the more intuitive 2D interpretation.

 Note that our Fig.~\ref{fig5}(a) should not be understood in the sense  that the binodal
  region splits into a disjoint union of two components. In the actual
  four-dimensional phase space the binodal region is connected. It looks like
  a four-dimensional torus, whose three-dimensional cross-section is the
  body of revolution of the shaded region, around the bisector of the first
  quadrant. Each point $\BF_{+}$ on the binodal (the boundary of the binodal
  region) can coexist with a unique counterpart
  $\BF_{-}$ on the binodal. Each pair $\BF_{\pm}$ of such coexisting states is
represented by two pairs of points ($A$,$B$) and ($A'$,$B'$) in our phase
diagram Fig.~\ref{fig5}(a).  

More formally, viewed from the full 4D space of deformation
  gradients $\BF$,  these points represent traces of a  2D critical torus intersecting the subspace of diagonal matrices. Indeed, recall that a $2\times 2$ matrix $\BF$ can be written as $\BR\BD\BU$, where $\{\BR,\BU\}\subset SO(2)$ and $\BD$ is diagonal with
    nonnegative entries. If we represent each $\BF$ by its diagonal form
    $\BD$, then in the $2$-dimensional space of diagonal matrices each point
    corresponds to the entire $SO(2)\times SO(2)=S^1 \times S^1$ manifold in the
    $\BF$-space.  To express it more  generally,  each point in the $n$-dimensional space of singular values represents $n(n-1)$-dimensional manifold $SO(n)\times SO(n)$; the critical set by itself  is $n^2-2$ dimensional.

 In striking contrast to what we have seen in the case of
  'elastic liquids', the binodal region does not partition the phase space
  into disjoint phases, and the values of $\BF$ represented by the dashed line
  in Fig.~\ref{fig5}(a) correspond to homogeneous (affine)
 configurations which remain globally stable, as one travels from the high density
 phase to the lower. Even though   the   energy  in this domain   remains nonconvex,  the equilibrium system does not form  mixtures (microstructures)  due to the prohibitively high rigidity-induced extensive cost of mixing.
 
Finally, we observe that the ensuing optimal microstructure is layered (simple
laminate) with the layer normal being the
  common singular vector of $\BF_{\pm}$ (eigenvector of $\BC_{\pm}$)
  corresponding to singular values $\Ge_{\pm}$. 
    Since nonlinear elasticity is a
  scale-free theory,  such  microstructure does not have a scale and is
   represented only by  the values $\BF_{\pm}$ of
    deformation gradient in coexisting phases and the volume fraction
    controlled by the applied loading. More detailed information about the microstructure would be obtained in the models which incorporate an internal length scale which may be responsible, for instance,  for surface tension.
 
\subsection{Coherent  spinodal} 
 Equations \eqref{spineq1}, \eqref{spineq2} in the present setting have the form
  \begin{eqnarray}
\label{spineq1ex}
&&\mu\Ba_{0}+h''(d)d^{2}(\BF^{-1}\Ba_{0}\cdot\Bn_{0})\BF^{-T}\Bn_{0}=0,\\
\label{spineq2ex}
&&\mu\Bn_{0}+h''(d)d^{2}(\BF^{-1}\Ba_{0}\cdot\Bn_{0})\BF^{-1}\Ba_{0}=0.
  \end{eqnarray}
Eliminating unit vectors $\Bn_{0}$ and $\Ba_{0}$ from these equations we
obtain the characterization of the spinodal: $\BF\in\Md\CS$ if and only if
\begin{equation}
  \label{spinex}
  \mu\Ge_{\min}^{2}+d^{2}h''(d)=0,
\end{equation}
where $d=\det\BF$ and $\Ge_{\min}$ is the smallest singular value of $\BF$.
 While the   equation \eqref{spinex},  as well as   other related equations
 obtained below,  are valid in any number of space dimensions, we continue to illustrate them in
 two space dimensions. The spinodal, described by equation \eqref{spinex} is
 shown as a dashed line in Fig.~\ref{fig5}(a). We see that the spinodal lies
 entirely inside the binodal region, except for the critical set represented
 in Fig.~\ref{fig5}(a) by the  points $C$ and $C'$.

\subsection{Coherent  critical points} 
The exact location of the two (symmetry related) critical
points $C$ and $C'$ can be found as a solution of the
system of equations \eqref{spineq1},
\eqref{spineq2}, \eqref{cpe}.  For the Hadamard-Flory solid, equations \eqref{spineq1},
\eqref{spineq2} reduce to \eqref{spinex}, while equation \eqref{cpe} becomes
\[
h'''(d)d^{3}(\BF^{-1}\Ba_{0}\cdot\Bn_{0})^{3}=0.
\]
Taking a dot product of equation \eqref{spineq2ex} with $\Bn_{0}$ we obtain
$(\BF^{-1}\Ba_{0}\cdot\Bn_{0})^{2}d^{2}h''(d)=-\mu$, which implies that
$\BF^{-1}\Ba_{0}\cdot\Bn_{0}\not=0$. Thus, critical
points solve the system of equations \eqref{spinex},
 and
 \begin{equation}
   \label{crpteq}
h'''(d)=0.
 \end{equation}
For a typical double-well energy we expect that equation \eqref{crpteq} has a
unique root $d^{*}$. The two singular values
$\epsilon_{1}^{*}<\epsilon_{2}^{*}$ corresponding to the critical point are
then given by
\[
\epsilon_{1}^{*}=d^{*}\sqrt{-\frac{h''(d^{*})}{\mu}},\quad\epsilon_{2}^{*}=\frac{d^{*}}{\epsilon_{1}^{*}}.
\]
This implies that unless $\mu>-d^{*}h''(d^{*})$, there will be no critical
points on the spinodal, i.e., the solution of \eqref{spinex}, \eqref{crpteq}
will lie in the interior of the spinodal region and will be therefore unstable. Critical point
$(\epsilon_{1}^{*},\epsilon_{2}^{*})$ is shown as point $C$ in
Fig.~\ref{fig5}(a). The point $C'$, also shown in Fig.~\ref{fig5}(a)  is
related to $C$ by the coordinate interchange symmetry swapping $\Ge_{1}$ and $\Ge_{2}$.

 We reiterate that according to our general theory, these points are located  where the coherent spinodal and the coherent binodal touch each other in the
phase space of deformation gradients $\BF$.
In this perspective,  $C$ and $C'$ are not two
  different critical points and should be viewed instead as two visible (in
  our chosen section of the whole phase space)  representatives of the
  critical set. More formally, in the $n$-dimensional space of singular values the critical set 
  will be represented by $n^2-2 -n(n-1) = n-2$ dimensional submanifold. When
  $n=2$, the critical set will be zero-dimensional,  revealing itself  in such a representation,  as a discrete
  set of points.

 \subsection{Ground state energy} 
 While it can be proved that  the  equilibrium (ground state)  energy 
$\tilde W(\BF)$ in the model with nonzero rigidity cannot be presented in the form
$\tilde h(d)+\mu|\BF|^{2}/2$,
with $ h(d)$ replaced by some  new function  $\tilde h(d)$  \cite{grtrsolid},
 the 
explicit formula for $\tilde W(\BF)$ can still be obtained in the large rigidity limit.

To explain the construction, we focus on one of the two symmetric regimes, and set 
$\epsilon_{1}(\BF)\le\epsilon_{2}(\BF)$. We would also need to distinguish
between the supercritical ($\epsilon_{2}< \epsilon_{2}^*$) and subcritical
($\epsilon_{2}> \epsilon_{2}^*$) cases. 
\begin{figure*}
  \includegraphics[scale=0.18]{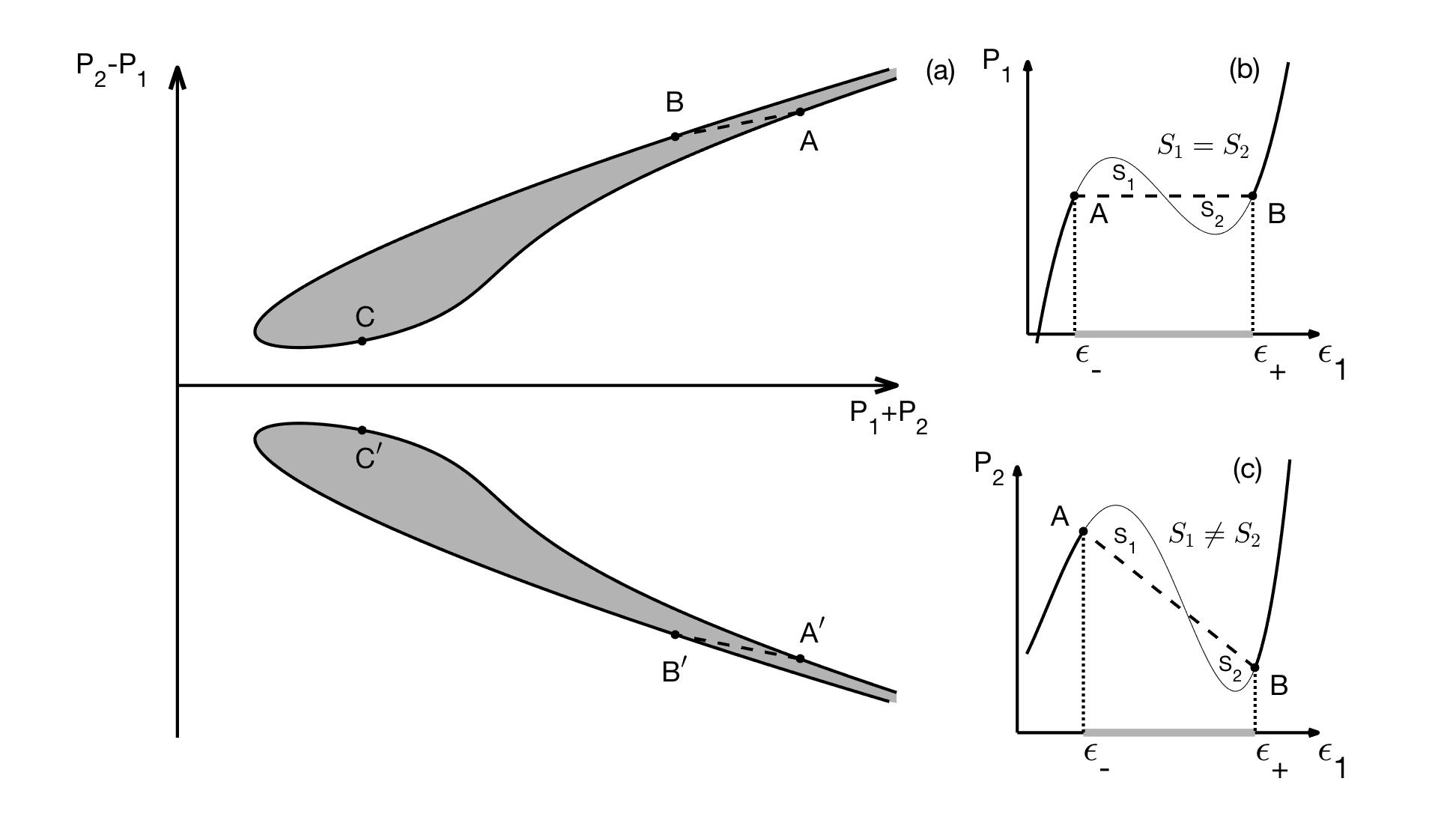}
  \caption{(a) Phase diagram in the space of principal stresses for an
    Hadamard-Flory solid with sufficiently large rigidity. (b,c) Two
    stress-strain relations in a rank one loading direction (hard device); the
    stress-strain correspondence across the binodal region is shown by the
    dashed lines.  Binodal region in (a) is shaded.  The analogs of the graphs
    (b) and (c) for the pair of points $A'$, $B'$ are identical to them except
    the horizontal axes would be $\epsilon_{2}$ and the labels $P_{1}$ and
    $P_{2}$ on the vertical axes would be interchanged.}
  \label{fig62}
 \end{figure*}

Recall that  since  in the supercritical case   mixing is suboptimal, we have $$\tilde W(\epsilon_{1},\epsilon_{2})=  W(\epsilon_{1},\epsilon_{2}).$$ In the subcritical case the expression for the equilibrium energy depends on whether the point $(\epsilon_{1}(\BF), \epsilon_{2}(\BF))$ is inside or outside the binodal region. Outside the binodal region
 the homogeneous configurations are stable and  we again have $\tilde W(\epsilon_{1},\epsilon_{2})=  W(\epsilon_{1},\epsilon_{2})$. For the states inside  the binodal region 
   the relaxed value of the energy can be also found since we know  that there exists   a
  uniquely determined    pair of coexisting strains 
$\epsilon_{-}(\epsilon_{2})<\epsilon_{+}(\epsilon_{2})$,
 solving \eqref{js}, \eqref{maxwell} with $\epsilon_{0}=\epsilon_{2}$.
Using these values, which represent parameters of  energy
  minimizing simple  laminate, we can write an explicit expression for the relaxed
energy inside the binodal region in the form
$$\tilde W(\epsilon_{1},\epsilon_{2})=\frac{\epsilon_{1}-\epsilon_{-}}{\epsilon_{+}-\epsilon_{-}}W(\epsilon_{+},\epsilon_{2})+
\frac{\epsilon_{+}-\epsilon_{1}}{\epsilon_{+}-\epsilon_{-}}W(\epsilon_{-},\epsilon_{2}).$$
 Note that despite the  simple mixture  appearance of this formula, 
  the  relaxed energy is nonconvex. The reason is that     the optimal simple laminate) microstructure  is prestressed due to the nontrivial  
  interaction (non-additivity) effects  encoded in highly nonlinear equations
  \eqref{js}, \eqref{maxwell}.   Those interactions  enter the relaxed energy implicitly via
  the nonlinear functions $\Ge_{\pm}(\Ge_{2})$.  This is manifested, for instance,  by the failure of the conventional
  common tangent construction which differs from the common area construction illustrated  in  Fig.~\ref{fig5}(b).  In other words, although the relaxed energy is a ruled surface, it is \emph{flat}  only in specific rank one directions while it still remains concave in some other non-rank one directions.

  \subsection{Stress space} 
To emphasize the nontrivial nature of the phase coexistence in the presence of nonzero rigidity, it is instructive  to map the obtained  phase diagram from  the \emph{strain} space into the   \emph{stress} space.  This will allow us to characterize the same phase equilibria in the space     of intensive  variables akin to, say  pressure and temperature,  used in the classical thermodynamics. An   experience with other  systems exhibiting long range interactions suggests that  in such an 'intensive'  representation the   conventional \emph{curves} representing  'liquid-like' phase coexistence would  transform into  extended \emph{domains} representing  'solid-like' phase  coexistence \cite{guru17,campa2009statistical}.  As our Fig.~\ref{fig62}(a) shows, that is exactly what is happening.

To explain this figure, we first recall   
 that in the coordinate frame  where $\BF$ is diagonal, the Piola stress $\BP$ is also diagonal
 with components
\begin{equation} \label{P12}
 P_{1,2}=\mu\epsilon_{1,2}+h'(\epsilon_{1}\epsilon_{2})\epsilon_{2,1}.
 \end{equation}
The   typical graphs of $P_{1}(\epsilon_{1})$ and $P_{2}(\epsilon_{1})$ at a
given $\epsilon_{2}=\epsilon_{2}^0$ 
  are shown in Fig.~\ref{fig62}(b,c).
Note, in particular,  that while  the  principal stresses $P_{1}$  are the same in the
 two coexisting phases  $A$ and $B$,  the corresponding principal stresses $P_2$  in the
 same configurations are different. 
 
 Note  also that, rather remarkably,   the relation
 $P_{2}(\epsilon_{1})$ does not satisfy the Maxwell equal area condition which is  nevertheless  respected  by the   relation
 $P_{1}(\epsilon_{1})$ along the same path $AB$ in the strain space.   This
   markedly different behavior is due to the tensorial (anisotropic) nature of
   stress in solids. Along the loading path $AB$, i.e. along the line segment joining
   $\BF_{+}$ and $\BF_{-}$ in the phase space, relations \eqref{31} and
   \eqref{33} are exactly the Maxwell relations we see operative in
   Fig.~\ref{fig62}(b);  other components of stress, like the one shown in Fig.~\ref{fig62}(c), are not obligated by  \eqref{31},
  \eqref{33} to satisfy  the Maxwell relations.  We emphasize that, despite their  apparent differences,  Fig.~\ref{fig5}(c) also illustrate   exactly the  same
   strain-stress relation.  More specifically,  
     along the path $AB$ we have   $d=\Ge_{1}\Ge_{2}^{0}$. Hence, the volumetric
     stress-strain response is effectively described by the function  $P_{1}(\Ge_{1})$ and  Fig.~\ref{fig5}(c) 
     simply shows the function $P_{1}(\Ge_{1})-\mu\Ge_{1}$  in terms of $d$,
     while  the graph of the function $P_{1}(\Ge_{1})$ is shown in Fig.~\ref{fig62}(b). 
     Note also that our  Fig.~\ref{fig5}(b) presents the double-well part
     $\tilde{W}(\BF)-\mu|\BF|^{2}/2$ of the relaxed energy. Subtracting from
     the energy density a quadratic function of $\BF$
component  transforms  the Maxwell `common tangent line' into a `common tangent
   parabola' shown  in Fig.~\ref{fig5}(b)   and changes the horizontal Maxwell line in Fig.~\ref{fig62}(b)  into a slanted Maxwell line in Fig.~\ref{fig5}(c).   In fact, one can show that our Fig.~\ref{fig62} illustrates  a general phenomenon: the
   Maxwell property of the stress component $\BP\Bn\cdot\Ba$.
     
The whole set  of coexisting equilibrium stress components parametrized by
 $\epsilon_{2}^0$, which is shown in Fig.~\ref{fig62}(a), illustrates  the
 relaxed response of our two phase Hadamard-Flory solid. One can see  the anticipated   opening of a  2D coexistence domain  in the space
 of intensive variables which, as we have already mentioned,  is typical for systems with long range interactions \cite{guru17}. In our case such an opening is indicative of  the presence  of metastability and hysteresis in a soft loading device  but not in a hard loading device; such ensemble dependence of the equilibrium response is yet another characteristic property of systems with long range interactions \cite{guru17}.  Other examples of the same effect can be found  in \cite{baja15,grtrhard} dealing with   different types of elastic response. 
 
Finally, note that the analysis of the $AB$ path conducted above, can be extended by symmetry to the  $A'B'$ path also indicated  in Fig.~\ref{fig62} (a) inside the second   symmetric
coexistence domain  parametrized   by $\epsilon_{1}^0$. (See also 
  the second connected component of the  binodal region in Fig.~\ref{fig5}(a).)   In particular, for the $A'B'$ path, the  graphs of $P_{1,2}(\epsilon_{2})$ (at a given $\epsilon_{1}=\epsilon_{1}^0$) would be   identical to the  graphs of $P_{2,1}(\epsilon_{1})$ (at a given $\epsilon_{2}=\epsilon_{2}^0$) along the  $AB$ path  shown in Fig.~\ref{fig62}(b,c).

\section{5. Geometrically linear  Hadamard-Flory solid }
  Since  the \emph{geometrical nonlinearity} is usually neglected in problems
  involving bulk phases \cite{llkp86,chlw95}, it is natural to ask whether the
  rigidity-induced critical points survive   in  the  model of the same solid but now  relying exclusively on linear
  strains. 
  
Observe first, that there are no immediate mathematical reasons for a
    nonlinear energy, which  depends on $\BF$ only through a special combination
\begin{equation} \label{eps}
\bm{\epsilon}=(1/2)(\BF+\BF^{T}),
\end{equation}
 not to have stable critical
    points.  Note next that the assumptions that would formally justify geometric linearization, 
  would also automatically justify physical linearization, and then there would be no critical 
  points;   this  last conclusion is vacuously true in the geometrically linearized theory  of elastic phase transformations between two
 physically  linear-elastic phases, since such
 an energy has no spinodal region.

\begin{figure*}
  \includegraphics[scale=0.18]{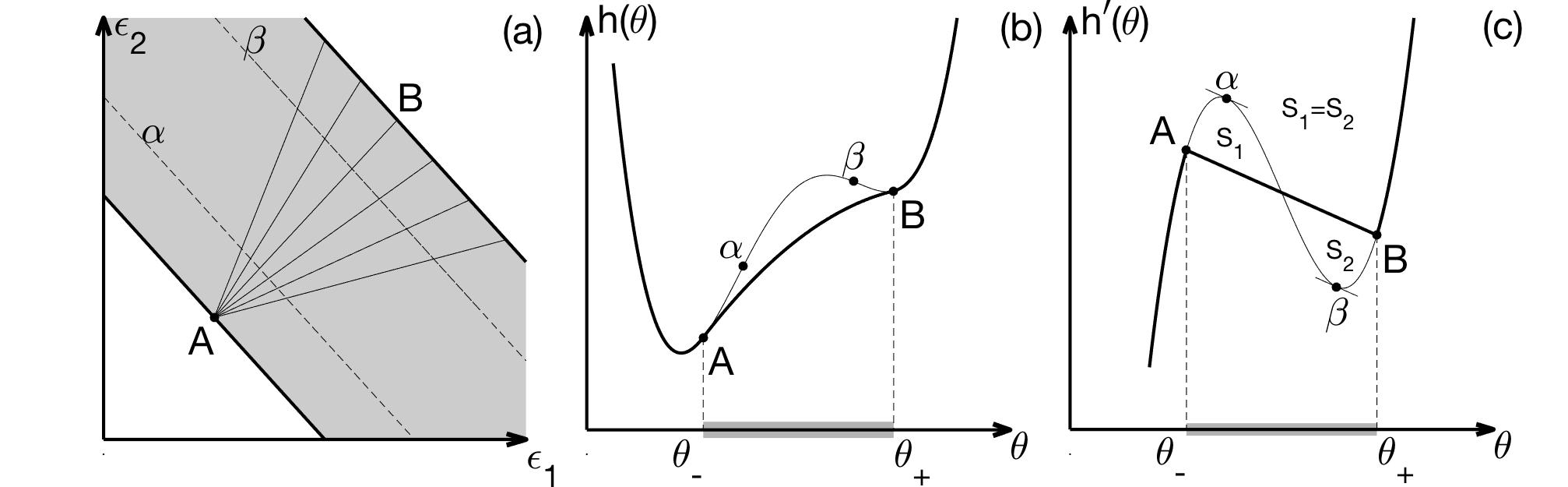} 
 \caption{(a) Phase diagram in the space of principal strains   for a geometrically linear   Hadamard solid. (b) 
Double-well   part of the  energy.  (c) The corresponding stress-strain relation.   The relaxed (ground state)  energy and the relaxed stress-strain response are shown by solid lines.  Binodal region in (a) is  shaded; dashed lines limit  the spinodal region.
 }
  \label{fig7}
 \end{figure*}  
  
 In general, one cannot
  expect to find stable critical points often. This is because critical points are points of
  tangency between the binodal and the spinodal,  and more
    often than not the spinodal lies entirely inside the binodal region. Even
when such tangency  does occur, stable
  critical points are in some sense least stable points on the binodal. For this
  reason,  any modification of the energy density function, making it easier to
  create energy-minimizing laminates, may destabilize parts of
  the binodal, in which case critical points might be the most
  vulnerable candidates for  destabilization.  This can be  already seen from the fact that if  
  one decreases the rigidity parameter $\mu$ in the Hadamard-Flory model
  \eqref{tr1}, the rigidity-induced critical points eventually lose their stability.

We emphasize, though, that it is not the geometrical nonlinearity, but the non-convex physical nonlinearity of the energy density  in $\BF$,
 which  is the main cause of instability leading to phase
 transition. Replacing in the energy density $\BF$ by $\bm{\epsilon}$ from \eqref{eps},  formally retains exactly the same type of non-convexity,  however it  makes the 
  destabilization of a homogeneous state and the formation of laminates qualitatively easier, since the pair $\BF_{\pm}$ is
  compatible if $\jump{\BF}$ is rank one, while the pair $ \bm{\epsilon}_{\pm}$ is
  compatible if $\jump{ \bm{\epsilon}}$ is rank two,  and some additional
  inequalities are satisfied. This means  that the compatibility set
  of $ \bm{\epsilon}_{\pm}$ has higher dimensionality than the compatibility set
  of $\BF_{\pm}$, which may in principle contribute to the destabilization of the   'bridge' between the stable components  of the coherent binodal, which we have seen forming in Hadamard-Flory solid  in the limit of sufficiently strong rigidity. Since the presence of such a  'bridge' is the factor, allowing the two phases to be smoothly connected, the stable critical points would then disappear as well.
  
As it follows from this discussion, the fate of coherent critical points under geometric linearization  is not clear  in the general case. However, as we show below,   the  rigidity-induced coherent critical points  do completely  disappear   if we replace our geometrically nonlinear Hadamard-Flory solid by its geometrically (but not physically) linear version.
In other words, critical points disappear if we consider a solid of Hadamard-Flory type, similarly  undergoing  a purely  volumetric phase transition, but now with
geometrically  linear but physically nonlinear elastic response in each of the phases. This example can be then  used as an   illustration of  the utmost  importance of geometrically exact description of elastic deformation in soft solids. 

In view of the  approximation $$\det\BF\approx 1+\mathrm{Tr}(\BF-\BI),$$ which is
valid in the limit $\BF\to\BI$,  the  natural geometrically linear analog of
the Hadamard-Flory energy density is
\begin{equation}
  \label{HFlin}
W(\BF)=h(\mathrm{Tr}\bm{\epsilon}) +\mu \vert\bm{\epsilon} \vert^{2}.
\end{equation}
where $\bm{\epsilon}$  is defined in  \eqref{eps}.
 Note that the geometrically linear version \eqref{HFlin} of
  the Hadamard-Flory solid is significantly less nonlinear than the
  original one. For example,
  for the double-well potential $h$, the geometrically nonlinear energy
  \eqref{tr1} is not rank-one convex, no matter how large $\mu>0$ is. By
  contrast, energy \eqref{HFlin} will become convex, when $\mu$ is
  sufficiently large, assuming the double-well potential $h$ is smooth.

\subsection{Coherent binodal}
It is natural to start  again with the characterization of  the  geometrically
linearized jump set, i.e., the   set   of equilibrium  coexisting strain tensors $ \bm{\epsilon}$. Using the kinematic compatibility condition  \eqref{jumpset1}  
which now takes the form $$\jump{{\bm{\epsilon}}}=(1/2)(\Ba\otimes\Bn+\Bn\otimes\Ba),$$ 
and the linearized analog of the traction continuity condition \eqref{31}
\begin{equation}
  \label{trcex}
   (\jump{h'(\mathrm{Tr}\bm{\epsilon})}\BI+2\mu\jump{\bm{\epsilon}}) \Bn=0,
\end{equation}
 we obtain 
$\Ba=\lambda\Bn$.
 Substituting $\Ba=\lambda\Bn$ back into \eqref{trcex}, we obtain
\begin{equation}
  \label{jtheta}
  \lambda=- \jump{h'}/2\mu.
\end{equation}
The
analog of condition   \eqref{32} 
 is now satisfied automatically, since  in geometrically linear elasticity the Piola stress tensor is  symmetric, and since for the linearized Hadamard-Flory solid, $\Ba$ is  a scalar multiple of $\Bn$. 

Finally, in the linearized theory, condition \eqref{33}  (generalized Maxwell  relation)
 takes the form $$\jump{h}-\lump{ h'}\jump{\theta}=0,$$ where $\Gth=\Trc\BGe$. This,
 together with \eqref{jtheta}, means geometrically, that
 $\theta_{-}=\Trc\BGe_{-}$ and $\theta_{+}=\Trc\BGe_{+}$  are the two points
 of common tangency to the graph of $ h(\theta)+\mu\theta^2$. Note however
 that, unless $\mu=0$,   this construction still differs from the
 conventional  Maxwell (common tangent)  construction for the volumetric part
 of the energy $h(\theta)$, see Fig.~\ref{fig7}(b,c). 
 Therefore, the
binodal region in the space $(\epsilon_1,\epsilon_2)$ of the eigenvalues of
the linearized strain tensor $\bm{\epsilon}$ is  delimited by  straight lines
$$\mathrm{Tr}\bm{\epsilon}=\theta_{\pm},$$ see Fig.~\ref{fig7}(a). Moreover, 
similar to the 'liquid' case and in contrast with the nonlinear 'solid'
case,  the optimal laminates in the geometrically linearized Hadamard-Flory theory are  not unique.  We
illustrate this effect  in Fig.~\ref{fig7}(a), by indicating schematically infinitely many  admissible
rank-one connections between the state $A$ and different states  $B$. 
It is also interesting that, in contradistinction with the geometrically exact theory, at large values of
$\mu$, the linearized binodal region may completely disappear as the energy becomes convex.

It is important to mention that, in contrast to what we had in geometrically nonlinear theory, here the morphology of nucleating precipitates in the infinite domain can be
  completely arbitrary, like in a
  liquid \cite{grtrmms}. In particular, for example, the microstructure may be represented by   simple laminates, as in geometrically nonlinear theory, but now  with an
  arbitrarily chosen orientation of layers. In view of the fact that the equilibrium configurations 
  do not satisfy our basic nondegeneracy condition $\BH>0$, the geometrically
  linear theory in this example exhibits extreme
  morphological non-uniqueness.  For instance, in  any homogeneously strained finite domain, Hashin's concentric sphere construction
  \cite{hash62} delivers infinitely many different (but energetically equivalent) minimizers
 with  fractal phase boundaries; the existence of such scale free minimizers is a general feature of elasticity theory, where the energy density depends only on $\Grad\By$.

  \subsection{Ground state energy} 
 
In this example  we can also compute the  relaxed (ground state) energy explicitly, which now can be written in the form \cite{grtrpcx}
  $$\tilde W(\BGe)=\tilde h(\Trc\BGe)+\mu|\bm{\epsilon}|^{2}.$$
 Inside the  binodal region, which is the domain of phase coexistence, $\Gth_{-}<\Trc\BGe<\Gth_{+}$, we find that
  \begin{equation}
    \label{hrel}
\tilde h(\theta)=\lump{ h'} (\theta-\lump{\theta})+\lump{ h} +\mu(\theta-\theta_{-})(\theta_{+}-\theta), 
\end{equation}
while $\tilde h(\theta)=h(\Gth)$, outside. Formula \eqref{hrel} can be
easily derived from the general relaxation formula
\cite[Lemma~4.3]{grtrnc}
\[
\Tld{W}(\BF)=\tau W(\BF_{+})+(1-\tau)W(\BF_{-}),
\]
provided, $\BF=\tau\BF_{+}+(1-\tau)\BF_{-}$, for some $\tau\in(0,1)$, where $\BF_{+}$ and $\BF_{-}$ are stable, rank-one related, and satisfy the so called  normality condition $\Trc(\jump{\BP}^{T}\jump{\BF})=0$. Note that the laminate construction behind formula \eqref{hrel} produces
relaxed nonlinear potential $\tilde{h}$, which is different from the straightforward convexification 
 of $h(\theta)$, because of the persistent  extensive mixing effects, see Fig.~\ref{fig7}(b).

 \subsection{Coherent spinodal} 
In the geometrically linearized theory the equations of spinodal \eqref{spineq1}--\eqref{spineq2} reduce to 
 \begin{equation} \label{lin1}
  h''(\theta) =-2\mu.
   \end{equation}  
 According to equation \eqref{jtheta},
  \[
\jump{\Gth}=\Trc(\jump{\BGe})=\Ba\cdot\Bn=\Gl=-\frac{\jump{h'}}{2\mu}.
  \]
 Applying the Lagrange theorem, we conclude that at least one solution of
  \eqref{lin1} must be in the interval $(\Gth_{-},\Gth_{+})$. In fact,
  for double-well potentials $h(\Gth)$, equation \eqref{lin1} has
  exactly two solutions $\Ga$ and $\Gb$, shown by dashed lines in
  Fig.~\ref{fig7}(a) and also marked in Fig.~\ref{fig7}(b,c). As is
  evident from Fig.~\ref{fig7}(a), the spinodal never touches the binodal.

\subsection{Coherent critical points } 

 In the geometrically linearized setting
the tensorial equation \eqref{cpe}, defining the rigidity-induced critical points,
turns into
\begin{equation} \label{lin2}
h'''(\Gth)=0.
\end{equation}
 The  two equations \eqref{lin1}, \eqref{lin2}  for  a single
unknown $\Gth$ are generically incompatible. When they are compatible (say, at a 
special value of $\mu$ for a given $h$), the binodal region in
Fig.~\ref{fig7}(a) collapses to a line, coinciding with the similarly collapsed spinodal
region. For all other values of $\mu$ the critical points are absent. 
 To summarize,  if the geometrically exact model of a  solid is replaced by the simplified  geometrically linearized description,  rigidity-induced coherent critical points may completely disappear.  This result highlights,   for instance,  the crucial importance of geometrically exact description of elastic deformation in living cells and tissues where the necessity  of the appropriate  reformulation of the early geometrically linear  approaches  have been long realized \cite{llsb23,shokef2012scaling,nardinocchi2013elastic,feng2016nonlinear,
 szachter2023nonlinear,grekas2021cells}.

\section{6. Conclusions}

In this paper we studied the   role of finite elastic rigidity in diffusionless solid-solid first order phase transitions.   Relying on the developed methods of coherent thermodynamics of elastic solids we  corroborated the general claim  that the  rigorous incorporation of rigidity into thermodynamic theory of phase transitions can lead not only to quantitative, but also to qualitative effects. 

We focused on the question of possible existence,  in the  coherent (or kinetic) phase diagrams,  of the unconventional  type of   rigidity-induced 'coherent critical points', which are fundamentally different from the conventional critical points, encountered  in rigidity-free (liquid) systems. To answer this question we had to  systematically develop a  general theory of  such critical points in physically and geometrically nonlinear elastic solids. In particular, we presented,  for the first time, the complete set of explicit equations, allowing one to locate   'coherent  critical points'   in the space of deformation gradients.

Our analysis relies on the assumption (of kinetic origin), that
elasticity is  an equilibrium property of solids.  It implies strain
compatibility, which prevents atoms from exchanging places,  and ultimately  brings into continuum theory effective long range interactions. Taking such elastic long-range interactions into consideration   requires geometrically exact description of non-hydrostatic deformation, which, for instance, destroys the additivity of the energy, producing extensive mixing effects, and  making the energy of a phase mixtures  sensitive  not only to  volume fractions of the coexisting phases, as is the case in liquid  systems, but also to the geometrical details of two-phase microstructures.   This  violates one of the foundations of the Gibbsian thermodynamics of phase transformations  and expectedly  brings about significant implications. 

To demonstrate  that  the emergence of the new type of critical points  is one of the rigidity-induced \emph{qualitative} effects, we systematically studied  the
loss of  stability of  a  homogeneous state, when the deformation
gradient crosses into the coherent
binodal region. In this case, the knowledge of the geometry of emerging  energy-minimizing two-phase
configurations is crucial for the determination of
stability limits.  
Along the way one needs to  fully characterize the set of pairs of
deformation gradients that could coexist at the coherent phase
boundary.

We showed that  if the complement
  of the coherent binodal region in the space of deformation
gradient is connected,  which  requires that the corresponding energy wells are not geometrically compatible, there exists a possibility to pass from one phase to another without the actual sharp phase transformation. The existence of such passages implies the presence of  rigidity-induced critical points, whose detailed quantitative characterization  in the framework of nonlinear elasticity  constitutes the main result of the paper. 

To illustrate the obtained general results, we applied  the developed theory to the
description of  a zero temperature equilibrium response of  an
isotropic solid  undergoing a purely volumetric,  first order phase
transition.   Similar transformations in  liquid  systems, such as
liquid-gas phase transitions, usually exhibit critical points in the
pressure-temperature space, which is a natural consequence of the fact
that phases have the same symmetry and can, in principle, be continuously transformed into one  another. To eliminate this type of (classical thermodynamic) criticality,  we have chosen our parameters in such a way that the  conventional  critical points do not exist.

 We then showed that when  rigidity  is sufficiently large,  the
 entire set of coexisting phases is stable, ensuring that specially
 oriented  simple   laminates  are sufficient
 to characterize  the   ground state  energy. We further
 observed  that in the regime of interest,   the complement
  of the coherent binodal region is indeed connected. As we have already mentioned, this signals the emergence of the new type of
non-classical (coherent)  critical points. 

Around such points, which we  can locate by examining stability
of solutions of our fundamental system of  algebraic   equations \eqref{spineq1},
\eqref{spineq2}, \eqref{cpe},   one can expect   anomalous  fluctuations and the
critical opalescence, which have indeed been observed in swelling gels 
\cite{tana78,tfsnss80,lita92}.   The corresponding `coherent' scaling
relations remain to be determined. The crucial insight here may be that 
phase  nucleation  in coherent systems   involves the configurational
`de-localization' in the strain space  which can   interfere  with  the
conventional  real-space divergence of the correlation length.

 To emphasize that the  geometrically exact modeling of a solid was absolutely essential
 for capturing the coherent critical points in our example, we
 considered it in
 juxtaposition with  a standard  description of the same volumetric
 phase transition within the framework of linearized kinematics. Perhaps unexpectedly for many, we
 discovered  that in this,  more broadly  accepted  setting of elasticity
 theory,  the  rigidity-induced coherent critical points  disappear.  This
 discovery highlights  the  importance of not only physical, but also
 geometrical  nonlinearity,  which is necessary for the accurate accounting of elastic effects in highly deformable solids. In this sense, our study  can be viewed as a cautionary tale, stressing  the    importance of finite strains in the thermodynamic  description  of phase transitions in soft matter.
 
Finally, we mention that the applicability of our results is  not limited to
volumetric phase transitions in isotropic gels, even though  we  used the
well-studied swelling transitions in gels as our main example. Instead, our
approach is  sufficiently general to be used  for the description of
arbitrary anisotropic nonlinear elastic solids undergoing  first order phase transitions
with arbitrary transformation strains. In this sense, our results can be
expected  to  have   implications  in  a broader range of research fields, from
living matter  to artificial bio-mimetic metamaterials, exhibiting
symmetry preserving phase transitions.

  \section{Acknowledgments.} 
  YG was supported by the
  National Science Foundation under Grant No. DMS-2305832.  LT was supported by the French  grant ANR-10-IDEX-0001-02 PSL.

\section{ References}
\def\cprime{$'$} \ifx \cedla \undefined \let \cedla = \c \fi\ifx \cyr
  \undefined \let \cyr = \relax \fi\ifx \cprime \undefined \def \cprime
  {$\mathsurround=0pt '$}\fi\ifx \prime \undefined \def \prime {'}
  \fi\def\Ya{Ya}

\end{document}